\documentclass[11pt]{article}
\usepackage{graphicx} 
\usepackage{amsmath}
\usepackage{xcolor}
\usepackage{comment}
\usepackage{tcolorbox}
\usepackage{algorithm}
\usepackage{algorithmic}
\usepackage{amsfonts}
\usepackage{authblk}

\title{Bayesian-Guided Diversity in Sequential Sampling for Recommender Systems}

\author{Hiba Bederina}

\author{Jill-Jênn Vie}

\affil{Soda Team, Inria Saclay, Palaiseau, France\\
}
\date{}

\begin{document}

\maketitle

\begin{abstract}
The challenge of balancing user relevance and content diversity in recommender systems is increasingly critical amid growing concerns about content homogeneity and reduced user engagement. In this work, we propose a novel framework that leverages a multi-objective, contextual sequential sampling strategy. Item selection is guided by Bayesian updates that dynamically adjust scores to optimize diversity. The reward formulation integrates multiple diversity metrics—including the log-determinant volume of a tuned similarity submatrix and ridge leverage scores—along with a diversity gain uncertainty term to address the exploration–exploitation trade-off. Both intra- and inter-batch diversity are modeled to promote serendipity and minimize redundancy. A dominance-based ranking procedure identifies Pareto-optimal item sets, enabling adaptive and balanced selections at each iteration. Experiments on a real-world dataset show that our approach significantly improves diversity without sacrificing relevance, demonstrating its potential to enhance user experience in large-scale recommendation settings. \\
\end{abstract}


\section{Introduction}
\label{sec:introduction}
With the vast amount of content available on the internet, users of various digital platforms face growing challenges in finding items that align with their preferences. Recommender Systems have been developed to enhance the user experience by reducing search time and accelerating discovery, enabling users to quickly access relevant and personalized content.

User satisfaction is central to fostering engagement and advancing the provider’s strategic goals---whether commercial, educational, or societal---while also addressing ethical concerns. These goals vary by context: commercial platforms may aim to boost revenue or conversion rates~\cite{mesas2020exploiting}, while public or educational systems may emphasize diversity, fairness, or serendipity to encourage discovery and long-term engagement~\cite{ziegler2005improving, cantador2015cross}. A common objective across domains is to match user preferences to maximize utility~\cite{ricci2010introduction}. However, these goals often conflict—for example, improving accuracy may reduce diversity, while promoting novelty might compromise short-term relevance. As a result, modern recommender systems increasingly adopt multi-objective optimization to balance competing criteria~\cite{zheng2022survey}.

While multi-criteria approaches are well-established in recommender systems, few have explicitly employed Pareto front techniques to address competing objectives~\cite{ribeiro2014multiobjective}. This work proposes a multi-objective framework that directly tackles the trade-off between item diversity and user relevance—two often conflicting goals—by identifying solutions that jointly optimize both. In this formulation, relevance is captured using interaction-based embeddings that model user preferences, whereas diversity is assessed through dedicated metrics that quantify variability in the feature space of selected items.

Different diversity metrics may overlap or exhibit positive or negative correlations~\cite{jost2006entropy}, leading to potentially conflicting interpretations. For instance, variance captures the spread of a distribution but does not account for proportional disparities as entropy-based measures like the Theil Index do. One way to address this issue is to incorporate contextual information, ensuring that the selected diversity metric aligns with the specific objectives of the application. To this end,  we introduce a \emph{Contextual Diversity-Aware Sequential Sampling} framework.

Since user context is incorporated into the sampling process, it is essential to prevent the agent from overly favoring items that are distant from the user's context, which could lead to overly clustered selections. Accordingly, two notions of diversity are considered in the reward computation: \textit{intra-batch} diversity (measuring diversity within the selected subset) and \textit{inter-batch} diversity (capturing the diversity between the selected items and the user's context).

Additionally, to incorporate serendipity—by encouraging exploration beyond already booked or frequently selected items and enabling unexpected or less popular recommendations—an uncertainty factor based on information gain is introduced. Inspired by the LinUCB framework, this factor multiplies item scores, injecting a distinct form of uncertainty that fosters more diverse and exploratory behavior.

Various evaluation metrics are used to assess the quality of item batches recommended to users. To evaluate diversity, both intra-batch and inter-batch measures were computed across rounds and compared across different recommendation policies. These metrics track how item diversity evolves over time, particularly by capturing the number of distinct items selected. In addition to diversity, the system's ability to align with user preferences was assessed using precision and recall for both liked and disliked items, ensuring that the recommender maintains relevance while promoting diversity. Finally, the number of pulls per item is tracked to evaluate the extent of exploration performed by each policy, offering insight into how effectively the system balances exploration and exploitation.

Section~\ref{sec:state of the art} presents the state of the art in recommender systems, diversity metrics, and multi-objective criteria. Section~\ref{subsect:similarity} compares various kernels for quantifying similarities and dissimilarities, while Section~\ref{subsect:diversityS} explores strategies for sampling diversified subsets of items based on these computed similarities. The proposed approach is introduced in Section~\ref{sec:approach}, which details a Contextual Diversity-aware Sequential Sampling framework, including the mathematical formulations for handling rewards and uncertainty. Section~\ref{sec:algorithm} presents the core algorithm and its step-by-step procedures. Section~\ref{sec:expres} provides an empirical evaluation of the proposed method using a well-known benchmark dataset, while Section~\ref{sec:discussion} analyzes its observed behavior and trade-offs. Finally, Section~\ref{sec:conclusion} summarizes the findings and outlines directions for future work.

\section{Related work}
\label{sec:state of the art}
Most recommender systems generally follow three main steps to suggest items to users: candidate generation, scoring, and ranking \cite{covington2016deep}. In the candidate generation step, representations of items are created within a shared latent space. These representations consist of fixed-size latent factors, with larger sizes incorporating more engineered features to capture richer patterns. These patterns can be derived from both user-item interactions and the intrinsic features of items. The generated representations ensure that similar items are positioned closer together in the latent space, aligning users with items that best suit their preferences \cite{he2017neural}.


The scoring step assigns a relevance score to each item, helping to select a subset that aligns with the user’s interests or the strategic objectives of the system \cite{he2017neural, shapira2022recommender}. After scoring, the ranking step identifies the most suitable items to suggest, with the goal of maximizing the company’s objectives, such as user satisfaction or profit \cite{covington2016deep}. By optimizing these steps, recommender systems aim to enhance the overall user experience while achieving commercial goals.

Recommender Systems can be broadly categorized into four types \cite{adomavicius2005toward}: Non personalized and stereotyped, collaborative filtering \cite{koren2021advances}, content-based \cite{javed2021review}, and Hybrid \cite{seth2022comparative,ccano2017hybrid} recommender systems. Non Personalized systems rely on simple statistics, ignoring individual user preferences and specific contexts, relying instead on aggregate statistics and general trends observed across the entire user base. While collaborative filtering leverages past user-item interactions, which can include implicit feedback (e.g., clicks, time spent on a page) or explicit feedback (e.g., ratings, likes, bookings). Content-based approaches, on the other hand, focus on the features of items and how they align with user preferences. Hybrid systems combine different approaches to overcome the limitations of individual methods and enhance recommendation quality \cite{adomavicius2005toward, seth2022comparative}.

Diversity has various interpretations and measures across different fields, such as biodiversity in ecology, income distributions in sociology, and diversity of styles or vocabularies in linguistics \cite{patil1982diversity}. A comprehensive survey on the definition and evaluation of diversity, as well as its impact on the quality of recommendations, is presented in \cite{kunaver2017diversity}. Different diversity metrics are used in the literature either to measure or evaluate the richness or evenness of a sample of elements. These metrics can be categorized into four main calculation methods: Entropy-based, Distance-based, Set-based, and Probabilistic/Distribution-based. Entropy-based metrics, like Simpson, Shannon \cite{cherednichenko2024generalizing} and Theil indices \cite{luo2023per} emphasizes the richness and evenness of a dataset. Distance-based metrics quantify pairwise similarities and are especially useful when coordinates or latent factors are available. Set-based metrics, also known as Coverage metrics, assess the extent to which recommendations span the item space. Rao's diversity \cite{rao1982diversity} (quadratic entropy) falls between these two categories, considering both pairwise dissimilarities and abundance. These metrics help measure diversity and fairness by assessing how much of the catalog base is utilized by the system. Special cases, like Long-tail Coverage and popularity Coverage, measure the skewness of recommendations towards the most or least popular items, thus avoiding bias. Besides, Ginin index \cite{tangirala2020evaluating} is a statistical measure used to evaluate inequality in items of recommendations, with higher Gini index values indicate a more unequal distribution of recommendations across items. Moreover Coverage \cite{ge2010beyond} is used to evaluate item space coverage, particularly for Top-K recommendations \cite{hiranandani2020optimization}, either for individual users or across multiple users
. Lastly, probabilistic measures like the Determinantal Point Processes (DPP) \cite{taskar2013determinantal} offer another approach to diversity measurement. For a deeper exploration of different evaluation metrics in recommender systems, particularly regarding company goals, we recommend referring to \cite{jadon2024comprehensive}.

Diversity is one of the beyond-accuracy criteria \cite{kaminskas2016diversity} that has been considered in the literature and categorized under the broader concept of item discoverability \cite{jannach2016recommendations}. \cite{kaminskas2016diversity} explores how optimizing each metric impacts others. Common beyond-accuracy measures in recommender systems include diversity, novelty, serendipity, catalog coverage, popularity bias \cite{abdollahpouri2017controlling}, and fairness \cite{ekstrand2022fairness}. For an extensive review of these criteria, we refer the reader to the survey by \cite{jannach2022multi}. In this paper, beyond-accuracy criteria are classified as one type of multi-objective criteria, termed Quality objectives, alongside Multistakeholder (strategic) Objectives \cite{anderson2020algorithmic} (e.g., relevance vs. profit), System-related Objectives \cite{steck2021deep} (e.g., accuracy vs. scalability), Time Horizon Objectives \cite{panniello2016impact} (e.g., customer retention, click-through rates), and User Experience Objectives \cite{nunes2017systematic} (e.g., transparency vs. cognitive effort).


Despite the extensive work on recommendation accuracy and fairness, few studies explicitly address how contextual diversity and exploration interact within sequential recommendation tasks. This work proposes a framework to bridge that gap.

\section{Diversity Quantification}
\label{sec:DivQuant}
Inspired by \cite{taskar2013determinantal}, which shares many similarities with Rao’s diversity \cite{rao1982diversity} (primarily used in Biodiversity and Ecology), we consider an application case where the kernel, \( L \), is expressed as a Gram matrix \( L = \phi^\top \phi \), where \( \phi \) contains the embeddings of the items. Each term 
for an item \( i \) is described as the product of a quality factor \( q_i \in \mathbb{R}^+ \) and its normalized embedding \( \phi_i \in \mathbb{R}^d \), with \( \lVert \phi_i \rVert = 1 \) (where \( d \) represents the number of features). The kernel entries can then be written as:


\[
L_{ij} = 
\underbrace{q_i}_{\text{Quality/}\atop\text{Diversity Score}}
\underbrace{\phi^{\top}_i \phi_j}_{\text{Similarity}}
\underbrace{q_j}_{\text{Quality/}\atop\text{Diversity Score}}
\]

\( \phi_i^{\top} \phi_j \in [-1, 1] \) serves as a similarity measure, with dissimilarity being proportional to diversity, between items \( i \) and \( j \), while \( q_i \in \mathbb{R}^+ \) (respectively $q_j$) can be interpreted as the probability that item \( i \) (respectively $j$) is likely to be interacted with by the user (\( P_i^u(\text{Interaction}) \)) \cite{liu2022determinantal}.
However, $q_i$ does truly represent in the proposed method the proportionality towards the similarity to the user $\frac{1}{\psi_u \phi_i}$ to reflect the relevance. The resulted overall formula considering relevance becomes: 

\[
L_{ij} = 
\underbrace{\frac{1}{\sqrt{\psi_u \phi_i}}}_{\text{Quality}\atop\text{Score}}
\underbrace{\phi^{\top}_i \phi_j}_{\text{Similarity}}
\underbrace{\frac{1}{\sqrt{\psi_u \phi_j}}}_{\text{Quality}\atop\text{ Score}}
\]



\subsection{Similarity measures}
\label{subsect:similarity}
The calculation of the diversity score begins with the computation of the similarity matrix between items. For this purpose, the RBF kernel, the linear kernel, and the cosine similarity were tested on a small example of a 2D item embeddings: \{Book1, Book2, Movie1, Opera1, Concert1, Movie2, Book3\}.

\begin{figure}[h]
    \centering
    \begin{minipage}[t]{0.3\textwidth}
        \centering
        \includegraphics[width=\textwidth]{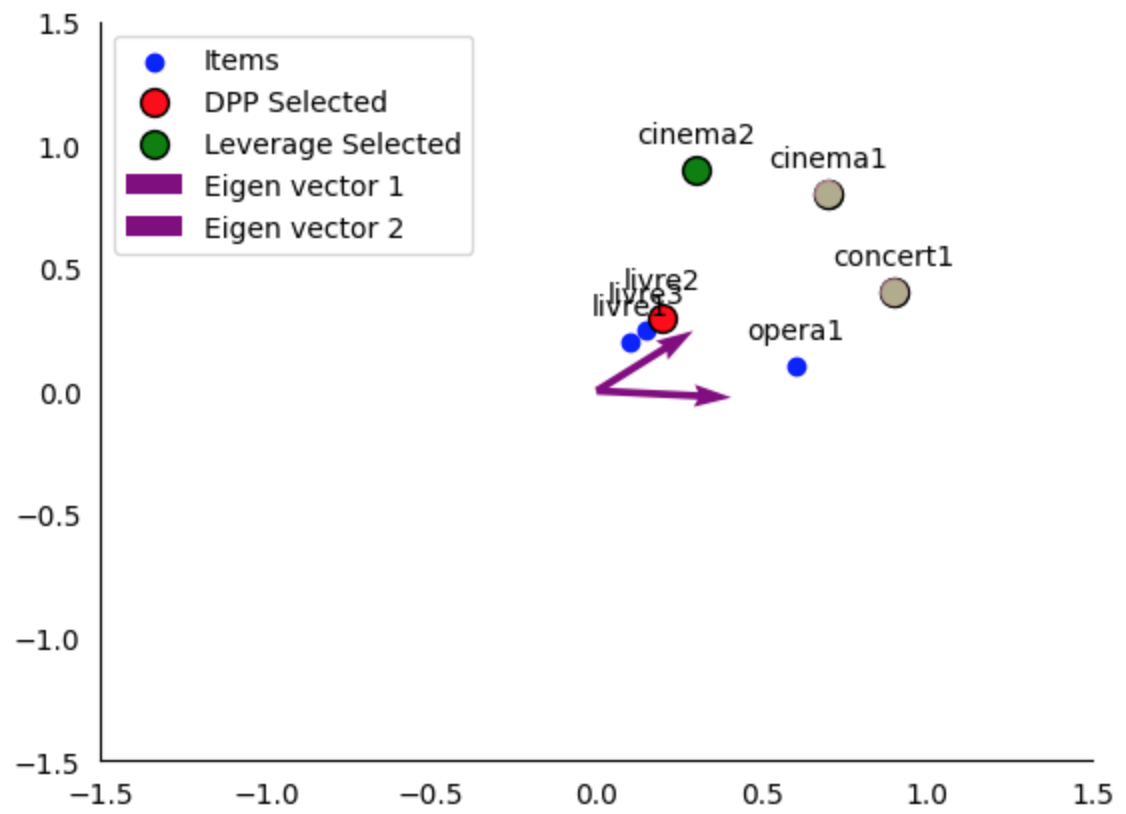}
        \caption{Linear Kernel}
        \label{lin_EV}
        \vspace{16pt}
        \[
            x^T y
        \]
    \end{minipage}\hfill
    \begin{minipage}[t]{0.3\textwidth}
        \centering
        \includegraphics[width=\textwidth]{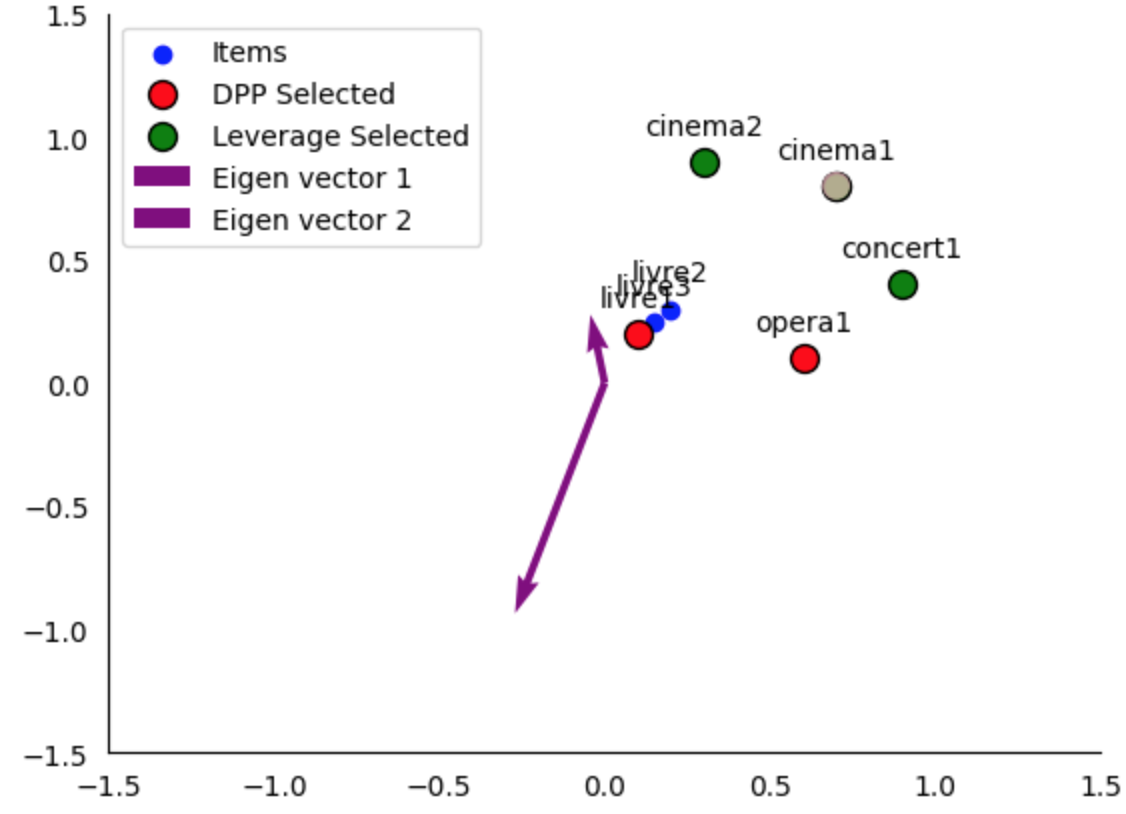}
        \caption{RBF Kernel}
        \label{rbf_EV}
        \vspace{11pt}
        \[
            \exp\left( -\frac{\left\| x_i - x_j \right\|^2}{2 \cdot \text{length\_scale}^2} \right)
        \]
    \end{minipage}\hfill
    \begin{minipage}[t]{0.3\textwidth}
        \centering
        \includegraphics[width=\textwidth]{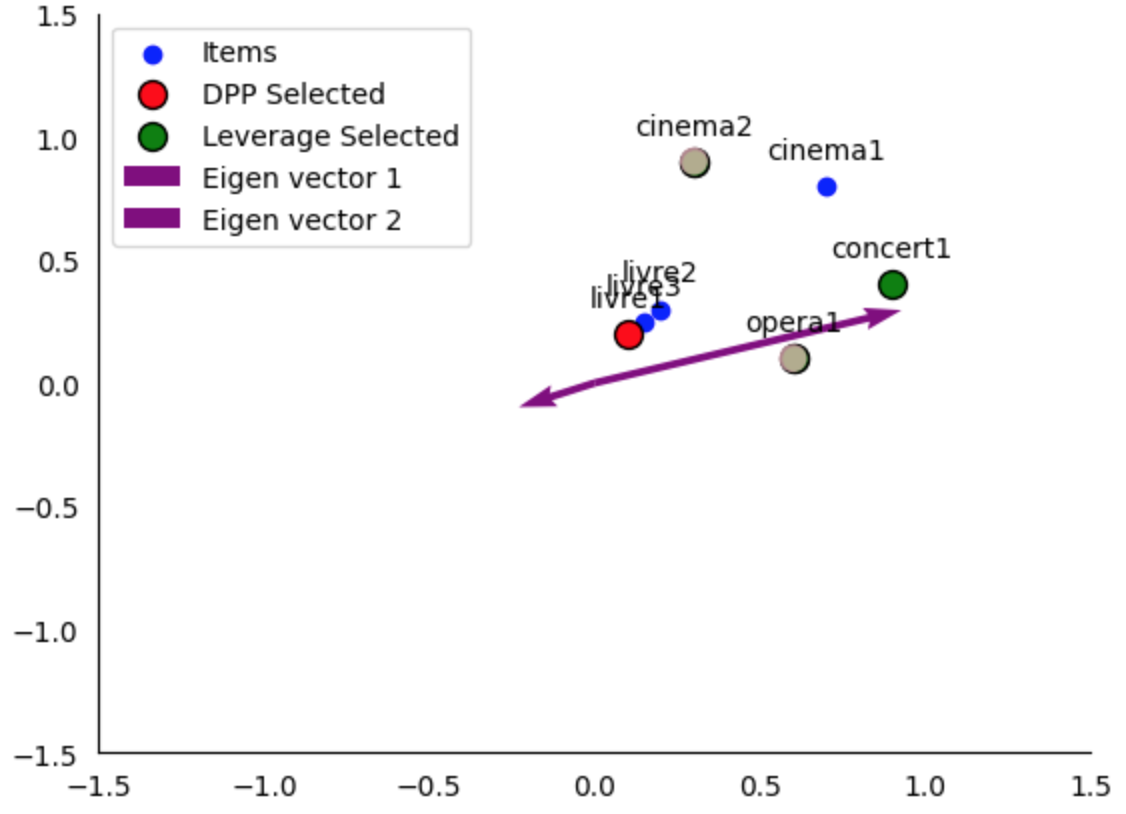}
        \caption{Cosine Similarity}
        \label{cos_EV}
        \vspace{2pt}
        \[
            \frac{\vec{A} \cdot \vec{B}}{\|\vec{A}\| \|\vec{B}\|}
        \]
    \end{minipage}
\end{figure}

Figures~\ref{lin_EV}, \ref{rbf_EV}, and~\ref{cos_EV} show that changing the kernel used for calculating similarities affects the orientation of the eigenvectors on the graph. The linear kernel appears to emphasize regions where items are most dispersed, when observing the projection onto the plane defined by the two eigenvectors corresponding to the largest eigenvalues. However, it tends to overlook items in the "book" category, even when they constitute a significant portion of the item set. 
The frequency of item categories may be considered when measuring  diversity. This difference between kernels may be due to a scaling issue.

\begin{figure}[h]
    \centering
    \begin{minipage}[t]{0.3\textwidth}
        \centering
        \includegraphics[width=\textwidth]{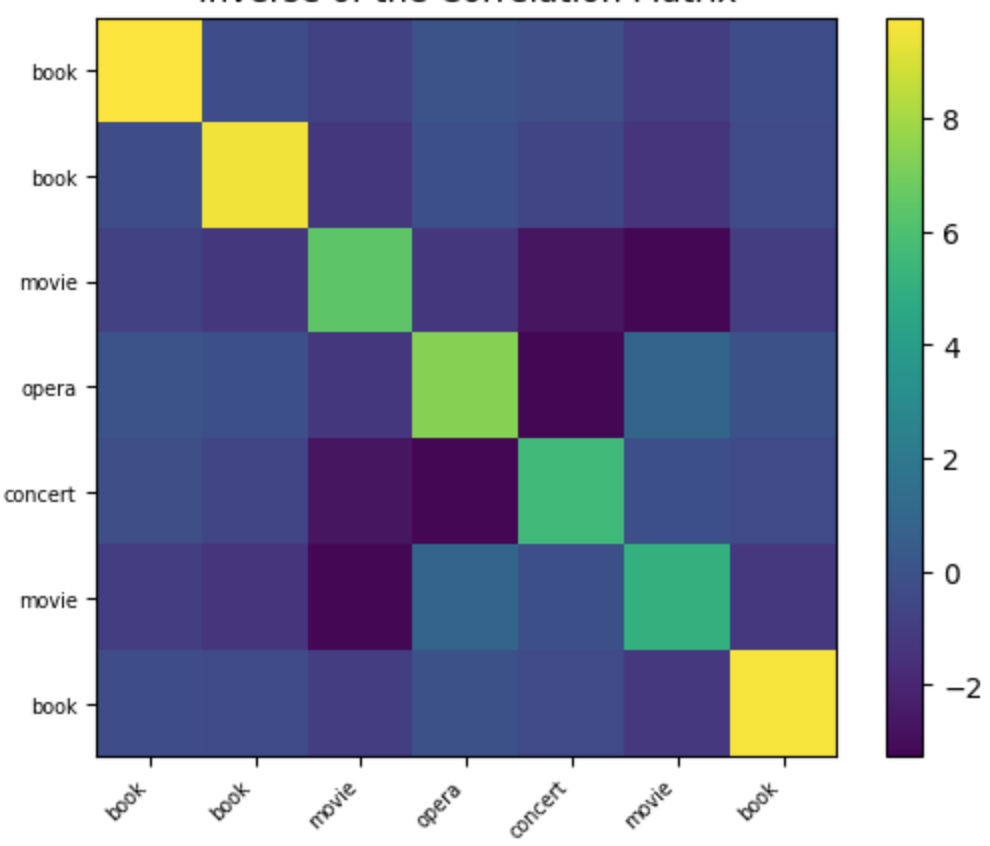}
        \caption{Linear Kernel}
        \label{lin_cov}
    \end{minipage}\hfill
    \begin{minipage}[t]{0.3\textwidth}
        \centering
        \includegraphics[width=\textwidth]{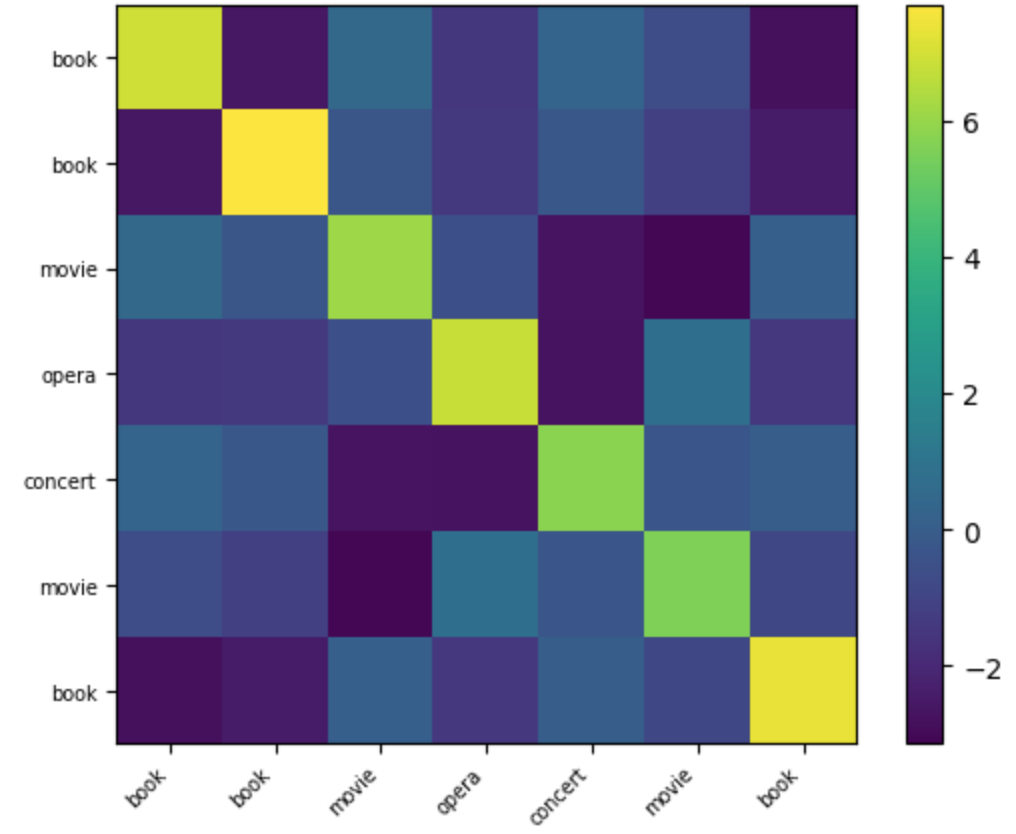}
        \caption{RBF Kernel}
        \label{rbf_cov}
    \end{minipage}\hfill
    \begin{minipage}[t]{0.3\textwidth}
        \centering
        \includegraphics[width=\textwidth]{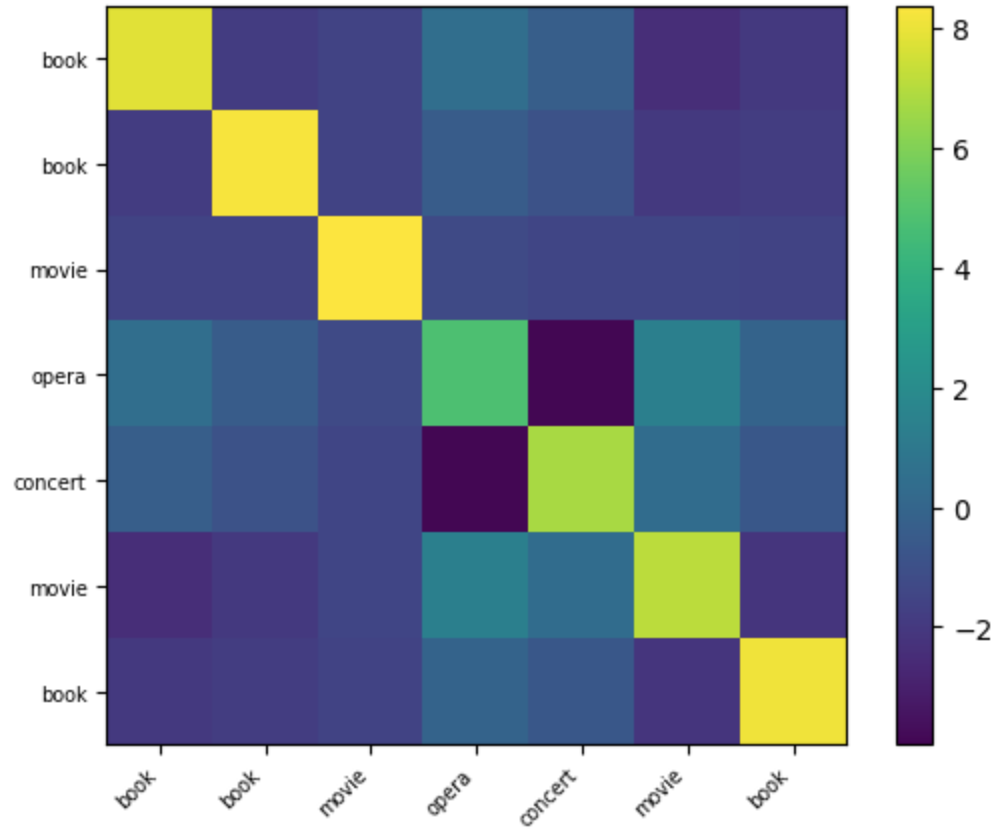}
        \caption{Cosine Similarity}
        \label{cos_cov}
    \end{minipage}
\end{figure}

Furthermore, Figures~\ref{lin_cov}, \ref{rbf_cov}, and~\ref{cos_cov} show that using the inverse of the covariance matrix is particularly effective in identifying items belonging to the same category. This is due to items within the same category tending to have similar values along the diagonal, which is visually reflected in the consistent coloring.

\subsection{Diversity metrics}
\label{subsect:diversityS}
Based on the specified kernel, two diversity metrics are examined in this paper:

\begin{itemize}
    \item Ridge Leverage Score: It is based on the inverse of the covariance matrix and quantifies the importance of each item based on the features.
    \item Volume: Measures the volume of the space occupied by the selected items (their spread). This is calculated as the product of the eigenvalues corresponding to each item.
\end{itemize}

The similarity matrix calculated for the textual embeddings of the items: "One Piece T. 11", "One Piece T. 12", "Huis Clos de Sartre", "One Piece T. 14", "One Piece: Film", "One Piece OST".

\begin{center}
\includegraphics[scale=0.3]{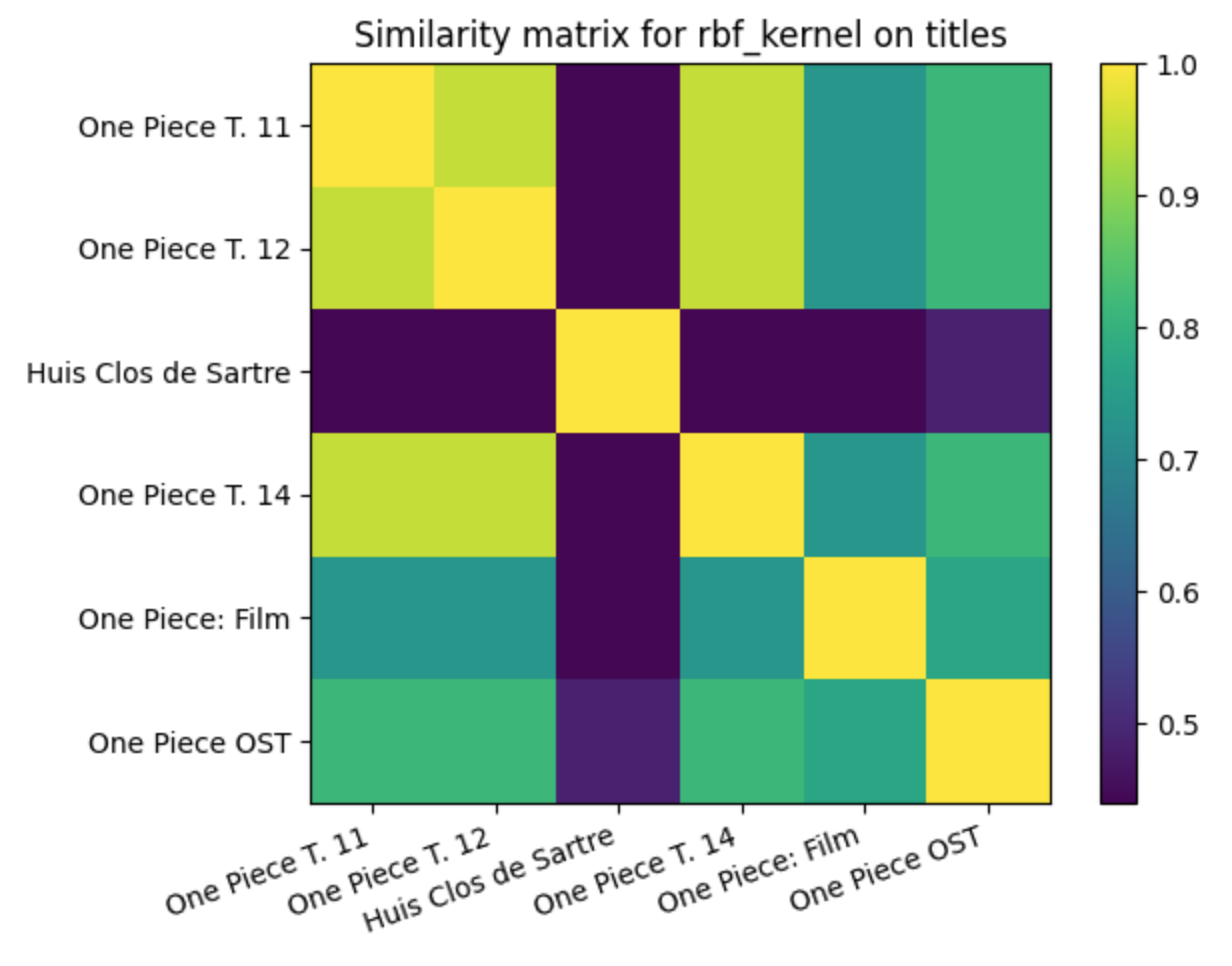}  
\end{center}

\begin{figure}[h]
    \centering
    \begin{minipage}[t]{0.45\textwidth}
        \centering
        \includegraphics[width=\textwidth]{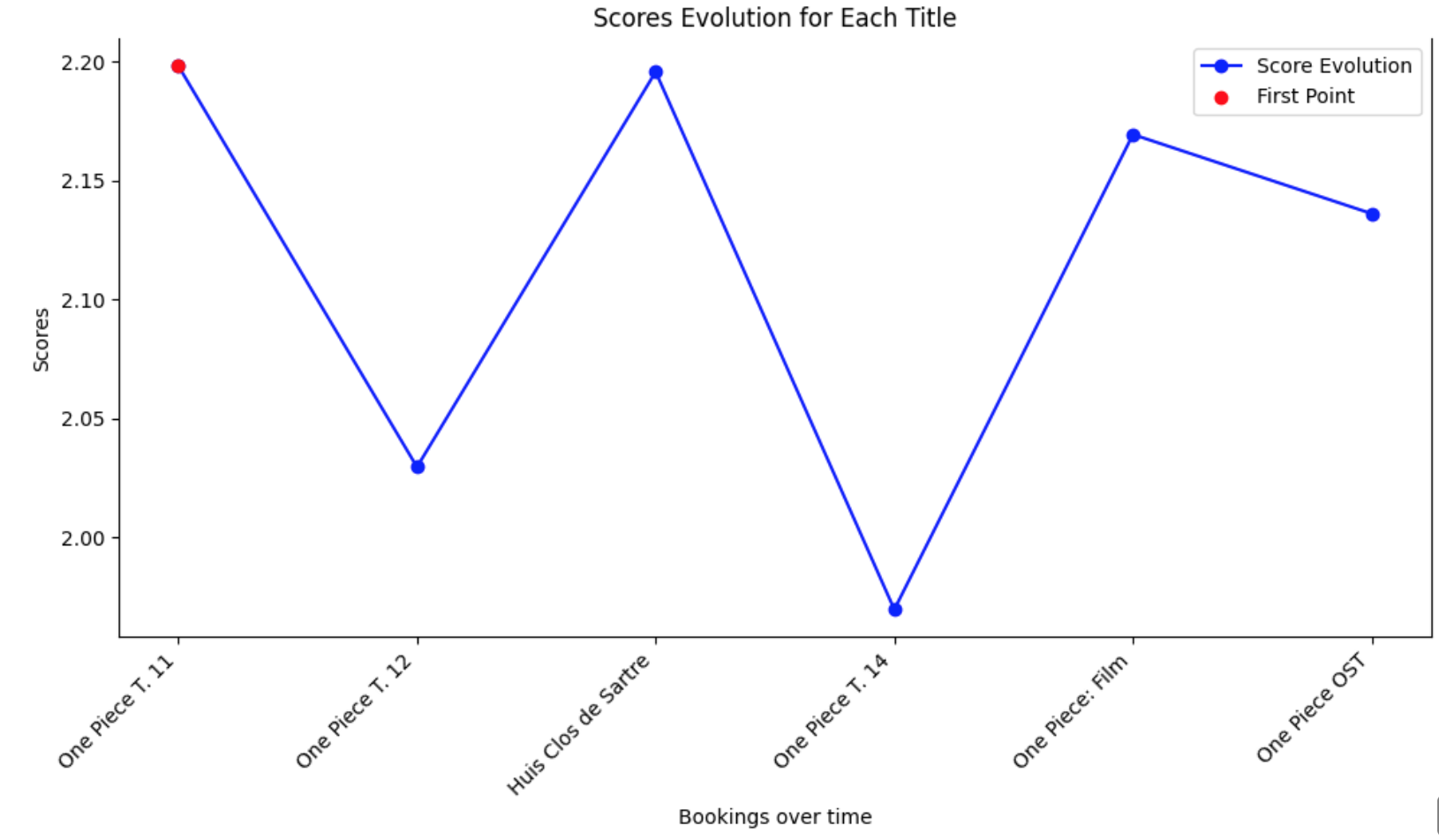}
        \\
        \caption{Ridge Leverage Score evolution}
        \label{RLSE}
    \end{minipage}\hfill
    \begin{minipage}[t]{0.45\textwidth}
        \centering
        \includegraphics[width=\textwidth]{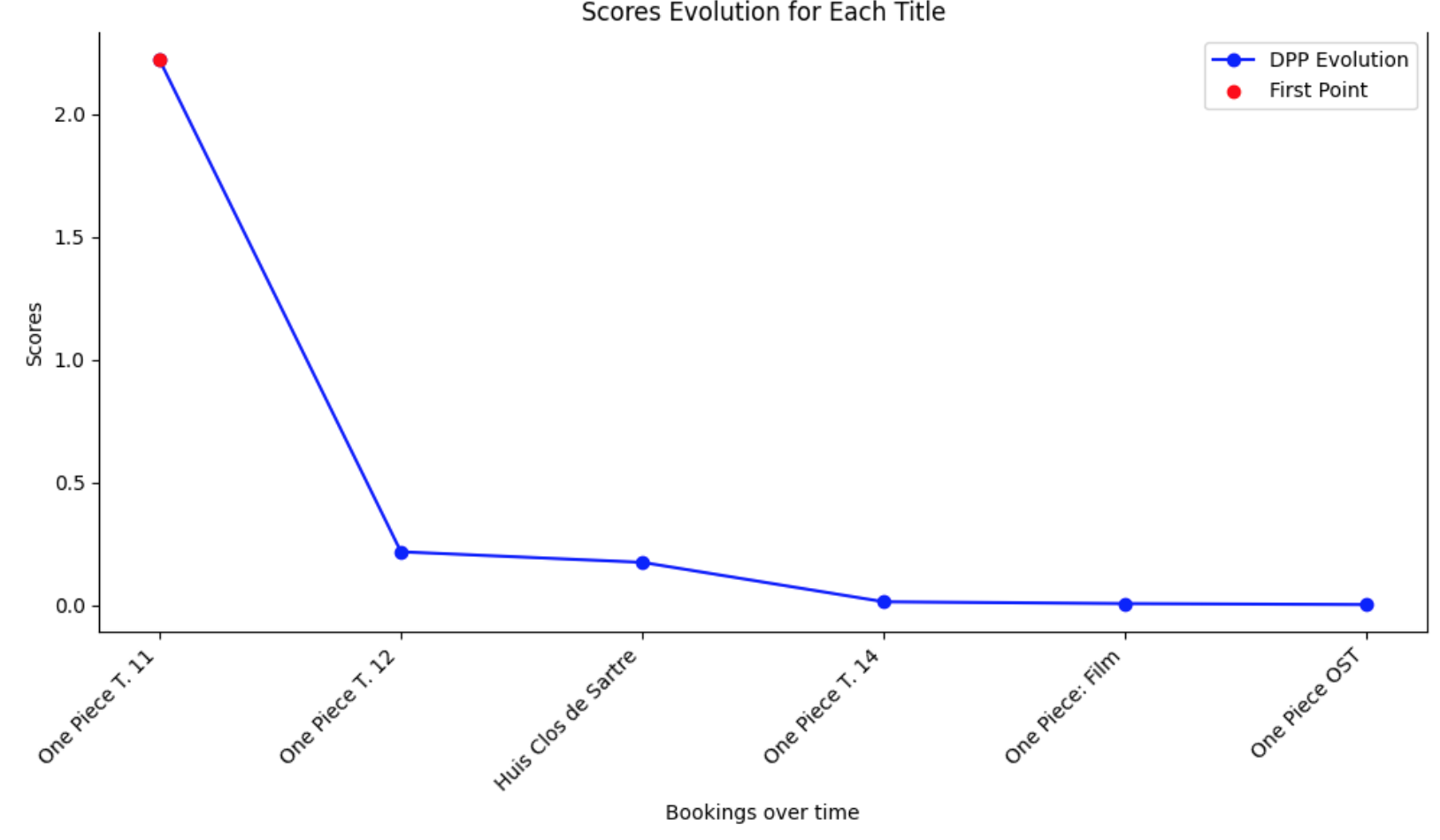}
        \caption{Volume}
        \label{volumeE}
    \end{minipage}\hfill
\end{figure}

Figures~\ref{RLSE} and~\ref{volumeE} illustrate how the Ridge Leverage Score and Volume metric evolve sequentially as items are booked.

The Volume can be used as a metric to evaluate \textit{intra-batch} diversity as the curve evolves cumulatively over time. On the contrary, Ridge Leverage Scores can be used to evaluate \textit{inter-batch} diversity as it is giving indication, not a quantification due to the range of values, about whether or not the newly added item is different from those in the history as the curve is strictly increasing or decreasing at each individual time step. 

\section{Contextual Diversity-aware Sequential Sampling}
\label{sec:approach}
We propose a \emph{Contextual Diversity-aware Sequential Sampling} approach. The method focuses on selecting a subset from a pool of \textit{items}—a core challenge in sequential learning scenarios. At each time step $t$, a new subset is selected. Each selection is informed by the user's current \textit{context}, which captures past bookings or interactions, and is guided by a \textit{reward} function aligned with the provider’s objective—\textit{diversity} in our case. Embeddings of both users and items serve as inputs to the decision process. The state corresponds to the user’s context or interaction history, while the actions represent the set of available items.

\subsection{Reward Function}
Two types of rewards are distinguishable: one for each item and another for the sampled subset of items. To evaluate the quality of the selected batch, two metrics are considered—Volume and the product (or sum of logs) of the ridge leverage scores RLS of the items within the subset. The Volume seeks to achieve the better \textit{intra-batch diversity} and the Ridge Leverage Score seeks to achieve better \textit{inter-batch diversity}. The goal is to enhance diversity with respect to the current context  by evaluating the Volume and RLS gains for the sampled subset at time step $t$. By leveraging the gain—i.e., the increment or submodular contribution  of each item to the batch diversity~\cite{gabillon2014large}—the approach prioritizes future selections and drives increases in both inter- and intra-batch diversity.

Let \(L \in \mathbb{R}^{n \times n}\) be a positive semi-definite similarity matrix, where each entry \(L_{ij}\) reflects the similarity between items \(i\) and \(j\).  
We denote by \(\mathcal{S} \subseteq I\) any subset of the overall set of items \(I = \{1, \dots, n\}\).  
Finally, let \( \mathcal{H} \in \mathbb{R}^{m \times d} \) denote the context matrix for all users, where \( m \) is the number of users.

\textbf{\textsc{Calculation of Volume: }}
Assuming \(\mathcal{S}\) is the selected batch, the submatrix \(L_\mathcal{S}\) is defined as
\(
L_\mathcal{S} = [L_{ij}]_{i,j \in \mathcal{S}}
\). The \textit{intra-batch diversity} of the subset \(\mathcal{S}\) is quantified by the determinant of \(L_\mathcal{S}\):
\[
\text{Volume}(\mathcal{S}) = \log\det(L_\mathcal{S}).
\]


\textbf{\textsc{Calculation of \(RLS\):}}
Let \(\{\kappa_i\}_{i=1}^{n}\) be a collection of positive real numbers measuring the quality of each item in terms of \textit{inter-batch diversity}. Let \(L_{\mathcal{H} \cup \{l\}} = [L_{ij}]_{i,j \in {\mathcal{H} \cup \{l\}}}\) be the kernel for items already in the history $\mathcal{H}$ together with the new item $l$. The Ridge Leverage Score of each item $l \in \mathcal{S}$ is computed as follows:

\[
\kappa_i = \left[ L_{\mathcal{H} \cup \{l\}} \left( L_{\mathcal{H} \cup \{l\}} + \lambda I \right)^{-1} \right]_{ii}
\]

The Ridge leverage score for the full batch \(\mathcal{S}\) is computed by summing the logarithm of each \(\kappa_i\), where $i \in \mathcal{S}$, which is often more numerically stable.

\[
\text{RLS}(\mathcal{S}) = \log\left(\prod_{i \in \mathcal{S}}^{} \kappa_i\right) = \sum_{i \in \mathcal{S}}^{} \log(\kappa_i)
\]

Let \(\Delta_{Volume}\) and \(\Delta_{RLS}\) denote the differences in Volume and Ridge Leverage Score  at time 
$t$ with respect to time $t-1$, respectively. This ensures consistency between the two gains, preventing them from offsetting each other. 

\[
reward^t = (\Delta_{Volume}) \cdot (\Delta_{RLS})
\]

\subsection{Exploitation-exploration dilemma} 
\label{subsection:expl}
Instead of directly adding an uncertainty bonus to the estimated reward, the approach draws on ideas from information theory by normalizing the gain by the number of times an arm has been selected. This is analogous to estimating the information gain—i.e., the item's potential to contribute to future diversity—based on the Thompson sample, while also accounting for its past selection behavior. By dividing by the count, the algorithm emphasizes not only the absolute uncertainty or potential gain, but also how informative or novel the selection of an item is.

\textit{Application to Thompson Sampling}: 
The actual reward at time step $t$ is used to update the Thompson sampling distribution for each selected item $i$. Specifically, the update rule modifies the parameters 
\(\alpha_i\) and 
\(\beta_i\) of the Beta posterior distribution. These posteriors, 
$\mathbb{P}(\alpha_i \mid reward^t)$  and $\mathbb{P}(\beta_i \mid reward^t)$, are updated based on the sign of the product of the two considered diversity metrics. The update rule for the parameters is as follows:

\[
\alpha_i \leftarrow \alpha_i + \max(0,reward^t),
\]
\[
\beta_i \leftarrow \beta_i + \max(0,-reward^t).
\]

That is, a positive reward at time \( t \) increases \( \alpha_i \), while a negative reward increases \( \beta_i \) by its absolute value.
This formulation can be expressed in terms of conditional probabilities as follows:
\[
\mathbb{P}(\alpha_i \mid reward^t) =
\begin{cases}
1, & \text{if } reward^t > 0, \\
0, & \text{otherwise,}
\end{cases}
\quad \text{and} \quad
\mathbb{P}(\beta_i \mid reward^t) =
\begin{cases}
1, & \text{if } reward^t < 0, \\
0, & \text{otherwise.}
\end{cases}
\]

At each time step $t$, a parameter \(\theta^{(t)}_i\) is sampled from the Beta distribution for each item \(i\):
\[
\theta^{(t)}_i \sim \operatorname{Beta}(\alpha_i, \beta_i).
\]

\textit{the Gain Ratio:}
Let \(\theta_i^{(t)}\) and \(\text{count}_i^{(t)}\) denote the cumulative value and the count for arm (or item) \(i\) at time \(t\), respectively. 
The gain ratio \(b_i^{(t)}\) from time \(t-1\) to \(t\) is given by:

\[
b_i^{(t)} = \frac{\theta_i^{(t)} - \theta_i^{(t-1)}}{\text{count}_i^{(t)}}.
\]

This formulation captures the change in the uncertainty estimate \(\theta_i\) for item \(i\) between two consecutive rounds, normalized by the number of times the item was selected at time \(t\). It effectively measures how the average contribution of each pull impacts \(\theta_i\) over time. A positive \(b_i^{(t)}\) indicates an increase in the average, while a negative value indicates a decrease.

\subsection{Scoring and Ranking Mechanisms}
Two objectives are considered, each weighted by an uncertainty value defined in Subsection~\ref{subsection:expl}: dissimilarity to the context (diversity) and similarity to the user (relevance). These objectives are incorporated  into the selection score computation through linear, nonlinear, or vectorized transformations. Multi-objective rankings rely on dominance-based selection to identify the items that contribute most to both diversification and relevance at time step $t$, treating the two objectives equally.

Let \( \mathbf{E}^I \in \mathbb{R}^{n \times d} \) denote the item embedding matrix, where \( n \) is the number of items 
; and \( b \in \mathbb{R}^{n \times 1} \) the weights vector.

\subsubsection{Linear Formulation}
The dissimilarity of the item under evaluation to the context is computed using the average distance, while its similarity to the user is also assessed. Both similarity and dissimilarity measures are computed using the linear kernel, which has been shown to perform best in sampling with DPP or when relying on Volume metric, as demonstrated in Section~\ref{subsect:similarity}. The term linear formulation refers to the combination of the considered objectives within a single linear expression. 
\\
\\
\textit{Objective 1:} The similarity of item \( i \) to the user is computed using the dot product:

\[
\text{sim}(\phi_i, \psi_u) = \phi_i^\top \psi_u
\]
\\
\textit{Objective 2:} The dissimilarity of item \( i \) to the context \( \mathcal{H}_u \) is defined as the average distance to user's history:





\[
\text{diss}(\phi_i, \mathcal{H}_u) = - \frac{1}{|\mathcal{H}_u|} \sum_{\phi_c \in \mathcal{H}_u} \phi_i^\top \phi_c
\]

This ensures that similarity to the user is maximized, while dissimilarity to the context is encouraged.
Then, to rank the items and select the top 
$K$, the sum of both objectives for each item $i$ is multiplied by the predicted  gain ratio for that item, denoted as $b_i$.

\[
b_i \times [\text{diss}(\phi_i, \mathcal{H}_u) + \text{sim}(\phi_i, \psi_u)]
\]

\subsubsection{Non Linear Formulation}
The nonlinear formulation relies on a distinct ranking strategy. It considers the same two previously described objectives but evaluates them jointly rather than aggregating them. To this end, instead of summing the objectives, we apply a \textit{dominance-based ranking} strategy:

\begin{itemize}
    \item An item \( i \) \textbf{dominates} an item \( j \) if it is at least as good as \( j \) in one objective and strictly better in another, when weighted by the gain ratio. Formally, item \( i \) dominates item \( j \) if:

    \[
    \left( b_i \times \text{sim}(\phi_i, u) \geq b_j \times \text{sim}(\phi_j, u) \right) \wedge \left( b_i \times \text{diss}(\phi_i, C) > b_j \times \text{diss}(\phi_j, C) \right)
    \]
    or
    \[
    \left( b_i \times \text{sim}(\phi_i, u) > b_j \times \text{sim}(\phi_j, u) \right) \wedge \left( b_i \times \text{diss}(\phi_i, C) \geq b_j \times \text{diss}(\phi_j, C) \right)
    \]
    
    \item Non-dominated items form the \textit{Pareto front}, representing optimal trade-offs between relevance and diversity.
    \item The ranking is established by iteratively identifying Pareto fronts: the first front (rank 0) contains the non-dominated items, while subsequent fronts are assigned increasing ranks based on their dominance level.
\end{itemize}

This approach ensures a balance between relevance and diversity without arbitrarily weighting the two objectives.

\subsubsection{Vectorized formulation}
Item scores are computed based on values proportional to their projections onto the eigenvectors corresponding to the best $K$ eigenvalues from the $K$-DPP. These eigenvectors are derived from two matrices: a weighted similarity between the user's context and the full item set, and a weighted similarity to the user. In both cases, the weights are given by the previously defined gain ratio.

\vspace{1em} 
\subsubsection*{Diversity Estimation} The first objective is to compute the vectorized element-wise multiplication of the item embeddings, weighted by the gain ratio.
The Diversity matrix computation proceeds in three steps, as follows:

\begin{enumerate}
    \item \textbf{Weighting the Item Embeddings}
     Let \( \mathbf{b} \in \mathbb{R}^n \) denote the gain-based weight vector, and let \( \operatorname{diag}(\mathbf{b}) \in \mathbb{R}^{n \times n} \) be the diagonal matrix with \( \mathbf{b} \) on its diagonal. The weighted item embedding matrix is obtained by scaling each row of \( \mathbf{E}^I \in \mathbb{R}^{n \times d} \) with its corresponding weight:
\[
\widetilde{\mathbf{E}}^I = \operatorname{diag}(\mathbf{b}) \, \mathbf{E}^I, \quad \widetilde{\mathbf{E}}^I \in \mathbb{R}^{n \times d}.
\]
    \item \textbf{Transpose of the Weighted Embeddings}
    We denote the transpose of the weighted item embeddings by:
\[
\left( \widetilde{\mathbf{E}}^I \right)^\top \in \mathbb{R}^{d \times n}.
\]
    \item \textbf{Diversity Matrix Computation}
    Let \( \mathbf{C} \in \mathbb{R}^{m \times d} \) denote the user context matrix. The diversity matrix is then computed as:
\[
\mathbf{D} = \mathbf{C} \left( \widetilde{\mathbf{E}}^I \right)^\top = \mathbf{C} \left( \operatorname{diag}(\mathbf{b}) \, \mathbf{E}^I \right)^\top, \mathbf{D} \in \mathbb{R}^{m \times n}.
\]
\end{enumerate}

\subsubsection*{Relevance Estimation} 
The Relevance matrix computation proceeds in two steps, as follows:

\begin{enumerate}
    \item \textbf{Weighted Item Embeddings}
We apply this weighting to the item embedding matrix \( \mathbf{E}^I \in \mathbb{R}^{k \times d} \) to obtain:
\[
\widetilde{\mathbf{E}}^I = \operatorname{diag}(\mathbf{b}) \, \mathbf{E}^I, \quad \widetilde{\mathbf{E}}^I \in \mathbb{R}^{k \times d}.
\]
    \item \textbf{Element-wise Multiplication with the User Embedding}
We compute the element-wise product between the weighted item embeddings and the user embedding:
\[
\mathbf{R} = \widetilde{\mathbf{E}}^I \odot \boldsymbol{\psi}_u,
\]
where \( \odot \) denotes the Hadamard (element-wise) product. Since \( \boldsymbol{\psi}_u \) is broadcasted across all \( k \) rows, the resulting matrix \( \mathbf{R} \) has the same shape:
\(\mathbf{R} \in \mathbb{R}^{k \times d}.\)

\end{enumerate}

Combining both steps, the final relevance matrix is given by:
\[
\mathbf{R} = \left( \operatorname{diag}(\mathbf{b}) \, \mathbf{E}^I \right) \odot \boldsymbol{\psi}_u.
\]
\subsubsection*{Computation of Item Scores using \(k\)-DPP}
$k$-DPP~\cite{kulesza2011k} is used to compute diversification probabilities for each item, based on vectorized objectives derived from either the diversity or relevance matrix. This serves as a scoring mechanism to ensure alignment between the selection scores and the reward, which is based on Volume. Each item is assigned two scores that reflect its diversity and relevance. At the end, dominance-based ranking is applied to select items that jointly achieve high values across both objectives.\\
\noindent Given a similarity kernel matrix \(L \in \mathbb{R}^{n \times n}\) and a desired sample size \(k \leq n\), the goal is to compute the selection probability \(\pi_j\) for each item $j$ under the $k$-DPP framework, as outlined in Algorithm~\ref{alg:kDPP_Proba}. This algorithm computes item-level probabilities that favor the selection of diverse or relevant items, based on the kernel $Q$ structure—where $Q \in \{D,R\}$—as encoded in the matrix $L$ and its eigenvalue decomposition.\\
\noindent The process begins with the eigenvalue decomposition of $L$, expressed as \(L = Q \Lambda Q^\top,\)
where \( \Lambda = \operatorname{diag}(\lambda_1, \lambda_2, \dots, \lambda_n) \) contains the eigenvalues, and \(
V = [v_1, v_2, \dots, v_n]\) is an orthonormal matrix whose columns are the corresponding eigenvectors. To ensure numerical stability, eigenvalues are clipped such that 
\(
\lambda_i \leftarrow \max(\lambda_i, 0), \quad \forall i.
\)\\
\noindent Next, a probability distribution over the eigenvectors is defined. For each eigenvalue \( \lambda_i \), the associated probability is computed as
\(
p_i = \frac{\lambda_i}{1 + \lambda_i},
\)
which is then normalized to ensure a valid probability distribution
\(
p_i \leftarrow \frac{p_i}{\sum_{j=1}^{n} p_j}.
\)
Using this distribution, \( k \) indices are sampled without replacement from the set \( \{1, \dots, n\} \), yielding a subset \( S \subset \{1, \dots, n\} \) of selected eigenvectors. The corresponding matrix \( V_S \in \mathbb{R}^{n \times k} \) is then constructed by collecting the eigenvectors indexed by \( S \);
\(
V_S = [v_i]_{i \in S}.
\)
For each item \( j \in \{1, \dots, n\} \), a selection probability is computed by summing the squared entries of the \( j \)th row of \( V_S \):
\(
\pi_j = \sum_{i \in S} \left(V_S(j,i)\right)^2.
\)  Finally, the resulting scores are normalized to form a proper probability distribution 
\(
\pi_j \leftarrow \frac{\pi_j}{\sum_{l=1}^{n} \pi_l}.
\)
\begin{algorithm}[H]
\caption{\(k\)-DPP Item Diversification Probabilities}
\label{alg:kDPP_Proba}
\begin{algorithmic}[1]
\REQUIRE A kernel matrix \(L \in \mathbb{R}^{n \times n}\), desired sample size \(k\) 
\ENSURE Item probabilities \(\pi \in \mathbb{R}^{n}\) 
\STATE \textbf{Symmetrize:} \(L \gets \frac{1}{2}(L + L^\top)\) \\
\STATE \textbf{Eigen Decomposition:} Compute \(L = Q \Lambda Q^\top\) \\
\FOR{\(i = 1,2,\dots,n\)}
    \STATE \(\lambda_i \gets \max(\lambda_i,0)\) \\
    \STATE \(p_i \gets \frac{\lambda_i}{1+\lambda_i}\) \\
\ENDFOR \\
\STATE Normalize \(p\) so that \(p_i \gets \frac{p_i}{\sum_{j=1}^{n} p_j}\) \\
\STATE \textbf{Sample:} Select a set \(S \subset \{1,\dots,n\}\) of size \(k\) without replacement according to \(\{p_i\}\) \\
\STATE Construct \(V_S \gets [q_i]_{i \in S} \in \mathbb{R}^{n \times k}\)
\FOR{\(j = 1,2,\dots,n\)}
    \STATE \(\pi_j \gets \sum_{i \in S} \left(V_S(j,i)\right)^2\)
\ENDFOR 
\STATE Normalize \(\pi\) so that \(\pi_j \gets \frac{\pi_j}{\sum_{l=1}^{n} \pi_l}\)
\RETURN \(\pi\) 
\end{algorithmic}
\end{algorithm}



\section{Overview of the Proposed Approach}
\label{sec:algorithm}
Algorithm \ref{alg:overall} describes the overall approach, which aims to sample a batch of \( K \) items for a user \( u \), maximizing both diversity and relevance in a sequential way setting over \( T \) rounds taking into account his history of seen or interacted with items. The input consists of the user embedding \( \psi_u \), item embeddings \( \mathbf{E}^I \), and sampling parameters. The output is a batch of items selected based on a reward function combining Volume and Ridge Leverage Score (RLS) gains.

The set of selected items for the user, \( \mathcal{H}_u \), is initialized to be empty. Each item \( i \in \{1, \dots, n\} \) is initialized with \( \text{count}_i = 0 \) to save how many times this item has been recommended, Thompson Sampling parameters \( \alpha_i = 1 \), \( \beta_i = 1 \), and an initial ratio gains \( \theta_i^{(t)} = 0 \) for all time steps \( t \in \{1, \dots, T\} \). Two variables, \( \text{Old}_{\text{Volume}} \) and \( \text{Old}_{\text{RLS}} \), are initialized to zero to track the reward changes across rounds.

At each round \( t = 1, \dots, T \), the algorithm calls the Batch Selection procedure defined in algorithm \ref{alg:pareto_selection} to return a candidate subset \( \mathcal{PF} \) of items balancing diversity and/or relevance—depending on the chosen policy—according to the current values of \( \theta, \alpha, \beta, \text{count} \), the user history \( \mathcal{H}_u \), \( \mathbf{E}^I \), and \( \psi_u \).

For each item \( i \in \mathcal{PF} \), the Ridge Leverage Score is computed with respect to the context (the selected subset) \( \mathcal{H}_u \), and the logarithm of each RLS value, 
is accumulated into \( \text{RLS}_\mathcal{PF} \). 

Once all items in the Batch have been evaluated, the algorithm computes the Volume score as the determinant of the kernel matrix modulated by the user embedding \( L_{\mathcal{PF}} \), denoted \( \text{Volume}_{\mathcal{PF}} \). The change in volume \( \Delta^{\text{Volume}}_{\mathcal{PF}} \) is computed relative to the previous round. Likewise, the change in RLS is given by \( \Delta^{\text{RLS}}_{\mathcal{PF}} = \text{RLS}_{\mathcal{PF}} - \text{Old}_{\text{RLS}} \).

The reward for the current round is defined as the product of the RLS and Volume gains:
\(
\text{reward}^t = \Delta^{\text{RLS}}_{\mathcal{PF}} \cdot \Delta^{\text{Volume}}_{\mathcal{PF}}.
\)
The Thompson Sampling parameters are subsequently updated based on the computed reward. For all items \( i \in \mathcal{PF} \), the updates are performed as follows:
if the reward is positive, then \( \alpha_i \gets \alpha_i + \mathbb{1}_{\{ \text{reward}^t > 0 \}} \); otherwise, \( \beta_i \gets \beta_i + \mathbb{1}_{\{ \text{reward}^t \leq 0 \}} \).
Here, \( \mathbb{1}_{\{\cdot\}} \) denotes the indicator function, which returns 1 when the condition is true, and 0 otherwise.

This process is repeated over \( T \) rounds, where at each step the system refines its selection by balancing exploration (serendipity) and the trade-off between relevance and diversity, thereby assisting the user with increasingly tailored suggestions each time a search is launched.

\begin{algorithm}
\caption{Contextual Diversity-aware sequential Sampling}\label{alg:overall}
\begin{algorithmic}[1]
\STATE \textbf{Input:} batch size $K$, user $u$, user embeddings $\psi$, item embeddings $\mathbf{E}^I$, 
\STATE \textbf{Output:} A batch of items achieving the highest rewards
\STATE Initialize $\mathcal{H}_u \gets \emptyset$ \COMMENT{Set of selected items}
\STATE Initialize \( \text{count}_i \gets 0 \quad \forall i \in \{1, \dots, n\} \).
\STATE Initialize \( \alpha_i \gets 1 \), \( \beta_i \gets 1 \quad \forall i \in \{1, \dots, n\} \)
\STATE Initialize \( \theta_i^{(t)} \gets 0 \quad \forall i \in \{1, \dots, n\},\; \forall t \in \{1, \dots, T\} \)
\STATE $Old_{Volume} \gets 0$
\STATE $Old_{RLS} \gets 0$
\FOR{$t = 1$ to $T$}
    \STATE $\mathcal{PF} \gets \text{Batch\_Selection}(K, t, \theta, \alpha, \beta, count, \mathcal{H}_u, \mathbf{E}^I, \psi_u)$ 

    \STATE $\text{RLS}_\mathcal{PF} \gets 0$
    \FORALL{$i \in \mathcal{PF}$}
        \STATE Compute $\text{RLS}_i \gets \text{Ridge\_Leverage\_Score}(i,\mathcal{H}_u)$
        \STATE Compute $\text{RLS}_\mathcal{PF} \gets \text{RLS}_\mathcal{PF} + \log(\text{RLS}_i)$
        
            \STATE $\mathcal{H}_u \gets \mathcal{H}_u \cup \{i\}$

    \ENDFOR
    \STATE Compute $\text{Volume}_\mathcal{PF} \gets \det \left(L_{\mathcal{PF}}\right)$
    \STATE $\Delta^{Volume}_\mathcal{PF} \gets \text{Volume}_\mathcal{PF} - Old_{Volume}$
    \STATE $Old_{Volume} \gets \text{Volume}_\mathcal{PF}$
    \STATE $\Delta^{RLS}_{\mathcal{PF}} \gets \text{RLS}_\mathcal{PF} - Old_{RLS}$
    \STATE $Old_{RLS} \gets \text{RLS}_\mathcal{PF}$
    \STATE Update $\text{reward}^t \gets \Delta^{RLS}_{\mathcal{PF}} * \Delta^{Volume}_\mathcal{PF}$
    \STATE Update \( \alpha_i \gets \alpha_i + \mathbb{1}_{\{\text{reward}^t > 0\}} \quad \forall i \in \mathcal{PF} \)
\STATE Update \( \beta_i \gets \beta_i + \mathbb{1}_{\{\text{reward}^t \leq 0\}} \quad \forall i \in \mathcal{PF} \)

\ENDFOR
\end{algorithmic}
\end{algorithm}

\paragraph{Batch Selection Procedure.}
Algorithm~\ref{alg:pareto_selection} selects a batch of items that jointly optimize relevance and diversity under various strategies, including mono-objective and multi-objective. It takes as input the batch size \( K \), the current time step \( t \), Thompson sampling parameters \( \theta, \alpha, \beta \), the item selection counts, the user's interaction history \( \mathcal{H}_u \), item embeddings \( \phi \), and the user embedding \( \psi_u \).

For each item \( i = 1, \dots, n \), the algorithm first samples a parameter \( \theta_i^{(t)} \) from a Beta distribution using Thompson Sampling based on \( \alpha_i \) and \( \beta_i \). It then computes a gain estimate \( b_i \) as the change in \( \theta_i \) between time steps \( t \) and \( t-1 \), normalized by the number of times the item has been selected.

Using this gain \( b_i \), a diversity score is computed as \( \text{div}_i = b_i \cdot \text{diss}(\phi_i, \mathcal{H}_u) \), where \( \text{diss}(\cdot) \) denotes the dissimilarity between the item's embedding and the user history. Similarly, a relevance score is computed as \( \text{rel}_i = b_i \cdot \text{sim}(\phi_i, \psi_u) \), where \( \text{sim}(\cdot) \) measures similarity to the user embedding.

Multiple scoring objectives are derived in the selection process. The aggregated score for each item \( i \) is defined as 
\( \text{Agg\_obj}_i = b_i \cdot (1 + \text{div}_i - \text{rel}_i) \), which combines both diversity and relevance signals. 
A single-objective relevance score is computed as 
\( \text{Sin\_obj}_i = b_i \cdot \text{rel}_i \). 
Additionally, a multi-objective formulation is considered as a tuple 
\( (b_i \cdot \text{rel}_i, \, b_i \cdot \text{div}_i) \), 
allowing both objectives to be treated simultaneously.


Next, to enable a vectorized treatment of both objectives, two kernel matrices are computed to obtain vectorized representations. The \emph{diversity matrix} is defined as 
\( \mathbf{D} = \mathcal{H}_u \left( \operatorname{diag}(\mathbf{b}) \, \mathbf{E}^I \right)^\top \), 
where \( \mathcal{H}_u \) denotes the user's history and \( \mathbf{E}^I \) the item embeddings. The \emph{relevance matrix} is given by 
\( \mathbf{R} = \left( \operatorname{diag}(\mathbf{b}) \, \mathbf{E}^I \right) \odot \boldsymbol{\psi}_u \), 
with \( \boldsymbol{\psi}_u \) representing the user embedding and \( \odot \) denoting the Hadamard (element-wise) product.
From these, \( k \)-DPP-based item selection probabilities are computed following the procedure detailed in Algorithm~\ref{alg:kDPP_Proba}:  
\(
\pi_{\text{div}} = \text{k\_DPP\_probabilities}(\mathbf{D}, K), \quad \pi_{\text{rel}} = \text{k\_DPP\_probabilities}(\mathbf{R}, K)
\).

Therefore, the returned batch of items is selected based on the specified strategy, which falls into one of two categories: dominance-based ranking or Top-$K$ selection.
\vspace{0.5cm}

\textit{Dominance-based ranking:}
\begin{itemize}
  \item \textbf{Vectorized Multi-Objective (VMo):} Items are ranked using dominance-based ranking over the vectorized scores \( \pi_{\text{div}} \) and \( \pi_{\text{rel}} \).
  \item \textbf{Non-Linear Multi-Objective (MO):} Dominance-based ranking is applied to the scalar scores \( \text{div}_i \) and \( \text{rel}_i \).
\end{itemize}

\textit{Top-$K$-based (argmax) ranking:}
\begin{itemize}
  \item \textbf{Aggregation (Lin):} Items are ranked and selected based on \( \text{Agg\_obj} \), which aggregates \( \text{div}_i \) and \( \text{rel}_i \).
  \item \textbf{Single Objective (SO):} Items are ranked by the relevance scores \( \text{rel}_i \).
  \item \textbf{Vectorized Mono-Objective (VMN):} Items are selected based on the vectorized diversity scores \( \pi_{\text{div}} \).
\end{itemize}

\begin{algorithm}[H]
\caption{Batch Selection}\label{alg:pareto_selection}
\begin{algorithmic}[1]
\STATE \textbf{Input:} $K$, $t$, $\theta$, $\alpha$, $\beta$, $count$, history $\mathcal{H}_u$, item embeddings $\mathbf{E}^I$, user embedding $\psi_u$
\STATE \textbf{Output:} Batch of size $K$
\FOR{$i = 1$ to $n$}
    \STATE $\theta_i^{(t)} \gets \text{ThompsonSampling}(\alpha_i, \beta_i)$
    \STATE $b_i \gets \frac{\theta_i^{(t)} -\theta_i^{(t-1)}}{\text{count}_i}$
    \STATE $\text{div}_i \gets b_i \cdot \text{diss}(\phi_i, \mathcal{H}_u)$
    \STATE $\text{rel}_i \gets b_i \cdot \text{sim}(\phi_i, \psi_u)$
    \STATE $Agg\_obj_i = b_i * (1 + div_i - rel_i)$
\ENDFOR
\STATE $\mathbf{D} = \mathcal{H}_u \left( \operatorname{diag}(\mathbf{b}) \, \mathbf{E}^I \right)^\top$
\STATE $\mathbf{R} = \left( \operatorname{diag}(\mathbf{b}) \, \mathbf{E}^I \right) \odot \boldsymbol{\psi}_u$
\STATE $\pi_{div} \gets \text{k-DPP\_probabilities}(D, K)$
\STATE $\pi_{rel} \gets \text{k-DPP\_probabilities}(R, K)$

\IF{way == "Vectorized Multi-Objective"}
    \STATE \textbf{return} $\text{Dominance\_ranking}(\pi_{div}, \pi_{rel})$
\ELSIF{way == "Multi-Objective"}
    \STATE \textbf{return} $\text{Dominance\_ranking}(div_i, rel_i)$
\ELSIF{way == "Aggregation"}
    \STATE \textbf{return} $\text{argsort(Agg\_obj)}$
\ELSIF{way == "Single objective"}
    \STATE \textbf{return} \text{argsort($\text{rel}_i$)}
\ELSIF{way == "Vectorized Mono-Objective"}
    \STATE \textbf{return} 
    \text{argsort($\pi_{div}$)}
\ENDIF

\end{algorithmic}
\end{algorithm}


\section{Experimental Results}
\label{sec:expres} 

All experiments were conducted in a browser-based environment using JupyterLite, which runs entirely client-side through WebAssembly via the Pyodide project. The computational environment used Python 3.12.7 (CPython), with the operating system reported as Emscripten 3.1.58 and architecture wasm32. Due to the nature of the Pyodide runtime, direct access to processor details was restricted.\\ 
All algorithms were implemented in Python using NumPy for efficient numerical computation, including Thompson Sampling through np.random.beta; pandas for loading and handling MovieLens dataset; scikit-learn for splitting the data into training and test sets as well as computing the linear kernel; scipy.sparse.linalg.svds for generating latent embeddings; DDPy for computing item diversity scores in the vectorized formulation; pymoo for dominance-based ranking using the NonDominatedSorting function to obtain the Pareto front; and matplotlib for visualizing the results.

\subsection{Case Study}



The MovieLens dataset is loaded as a table containing three columns: user ID, item ID, and the corresponding rating. To ensure reliable modeling, interactions are filtered to retain only items with more than five ratings. A custom stratified splitting strategy is then applied to divide the data into training and test sets while ensuring that all users and items are represented in both subsets. This procedure splits each user's ratings using \texttt{train\_test\_split} and reassigns a minimal number of interactions to guarantee that every item appearing in the training set also appears in the test set. After the split, users and items are reindexed to produce contiguous integer identifiers. The training set is then used to generate user and item embeddings via truncated Singular Value Decomposition (SVD), using \texttt{scipy.sparse.linalg.svds}. These latent factors capture historical user-item interaction patterns, placing similar items close to each other in the latent space, while user embeddings are positioned to reflect individual preferences. Finally, the proposed \emph{Contextual Diversity-aware Sequential Sampling} algorithm is applied to the test set using the learned embeddings to generate a personalized subset of recommendations for each user. Relevance is then assessed by verifying whether the recommended items recover user interactions from the test set that were not involved in the embedding construction.

\subsection{Implementation Details}

To measure the intra-batch diversity-based volume in high-dimensional space, we rely on the determinant of the kernel matrix, $\det(A)$. However, computing the logarithm of the determinant, $\log\det(A)$, is preferred for numerical stability and to leverage its submodular properties. Prior to this computation, the kernel matrix is verified and adjusted to ensure it is positive semi-definite. This involves symmetrizing the matrix by averaging it with its transpose, and regularizing it by clipping negative eigenvalues to zero after performing an eigen-decomposition using \texttt{numpy.linalg.eigh}. The matrix is then reconstructed from the filtered eigen-spectrum, and a small constant ($1 \times 10^{-6}$) is added to the diagonal to avoid near-singularities. The log-determinant is computed via \texttt{numpy.linalg.slogdet}, and the final volume is returned as $\exp(\log \det(K))$ if the sign is positive, or set to zero otherwise.

As part of the intra-batch diversity evaluation, variance was computed using \texttt{np.var}.

\subsection{Results analysis}
To comprehensively assess the performance of the proposed selection strategies, we conducted a series of evaluations targeting four key dimensions: diversity, relevance, exploration behavior, and learning dynamics. All results were averaged over 10 independent simulations to ensure robustness. Diversity was measured using statistical metrics such as volume, ridge leverage scores, and variance. 
Relevance was evaluated through precision and recall, both overall and separately for liked and disliked items. Exploration behavior was analyzed by tracking the distribution of item selections across rounds, revealing how widely each method samples the item space. Finally, learning performance was assessed via cumulative empirical regret in terms of Volume, capturing each strategy's ability to balance exploration and exploitation over time.\\

\begin{figure}[H]
    \centering
    \includegraphics[width=0.3\linewidth]{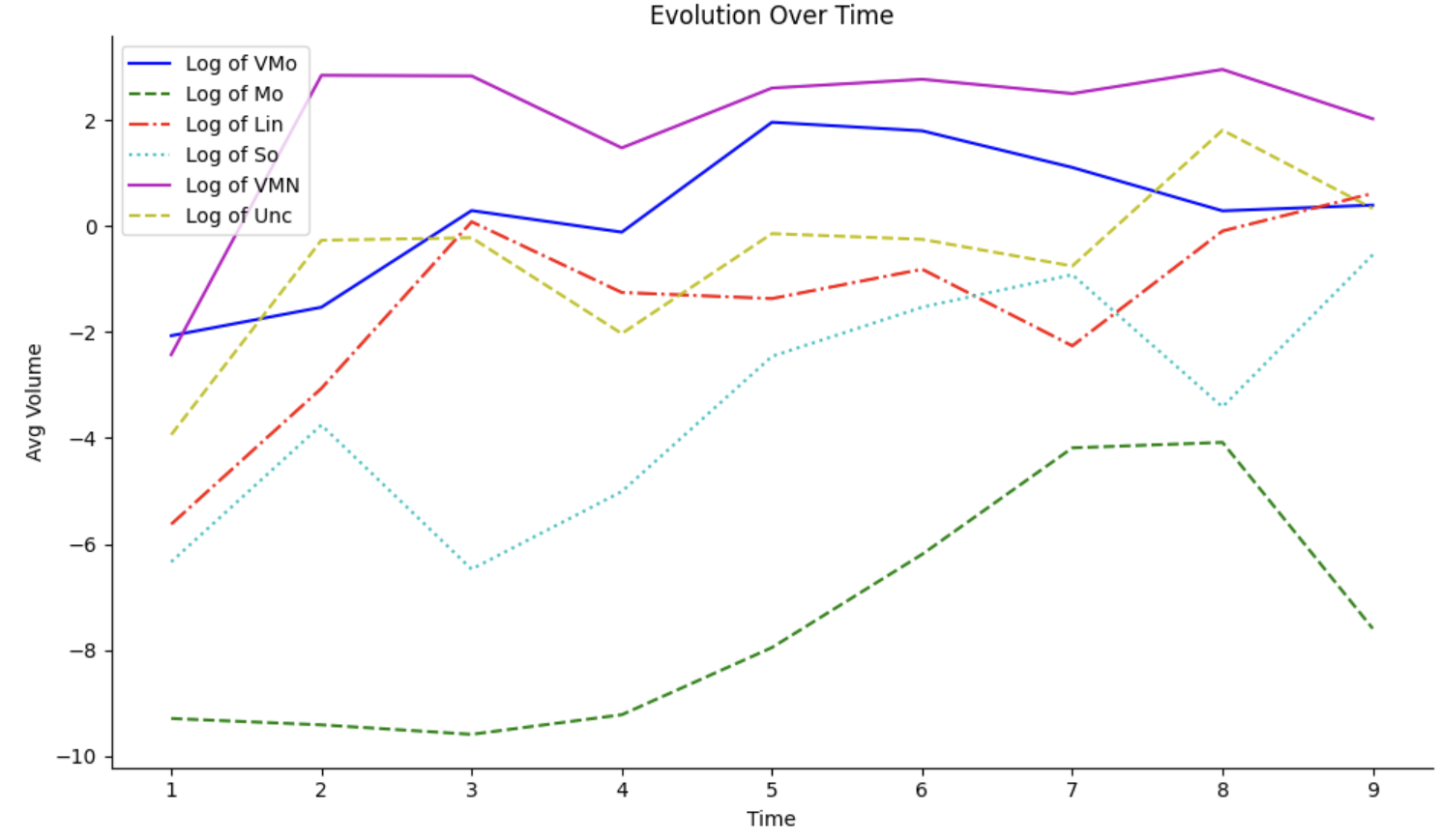}
    \includegraphics[width=0.3\linewidth]{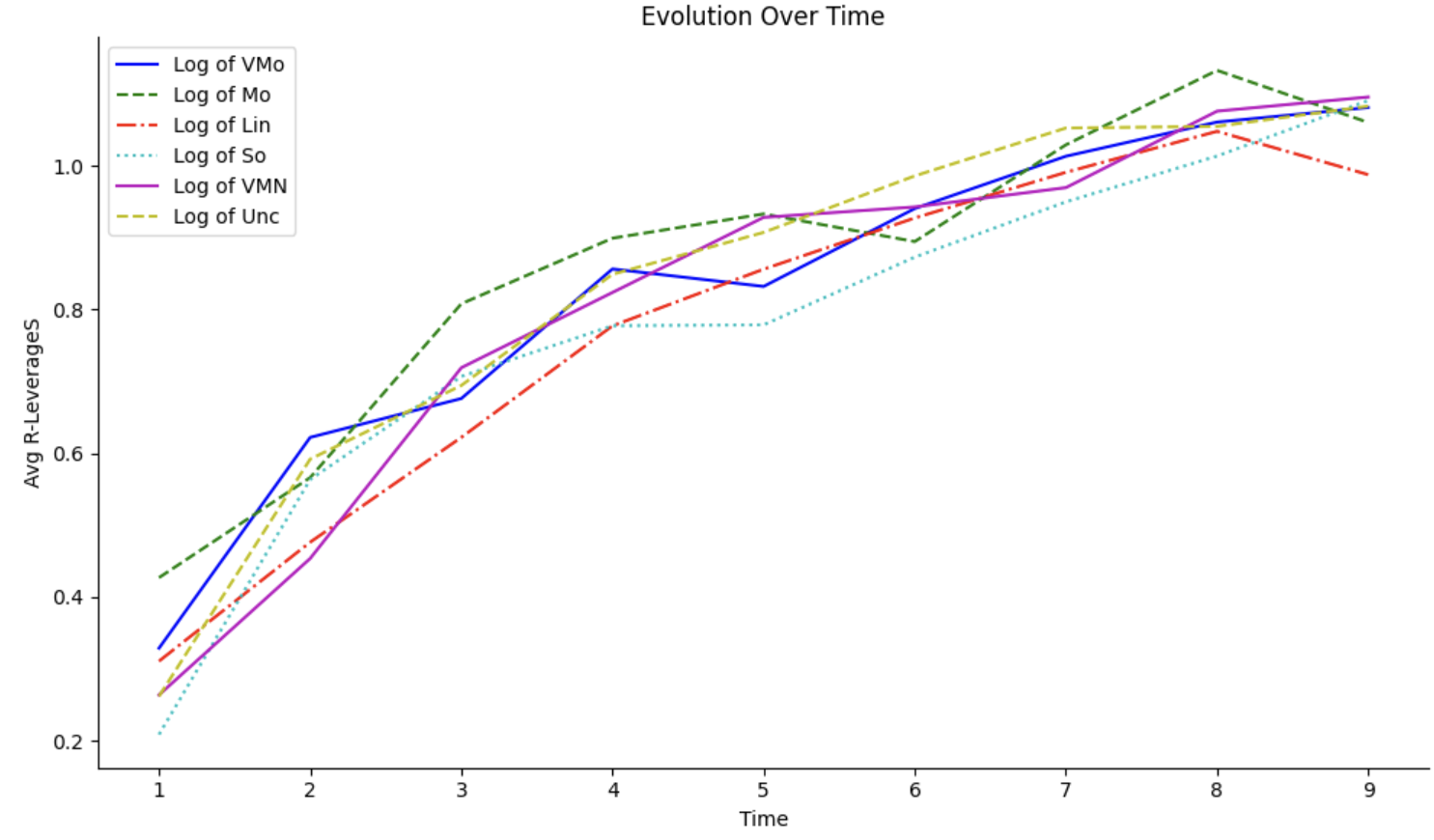}
    \includegraphics[width=0.3\linewidth]{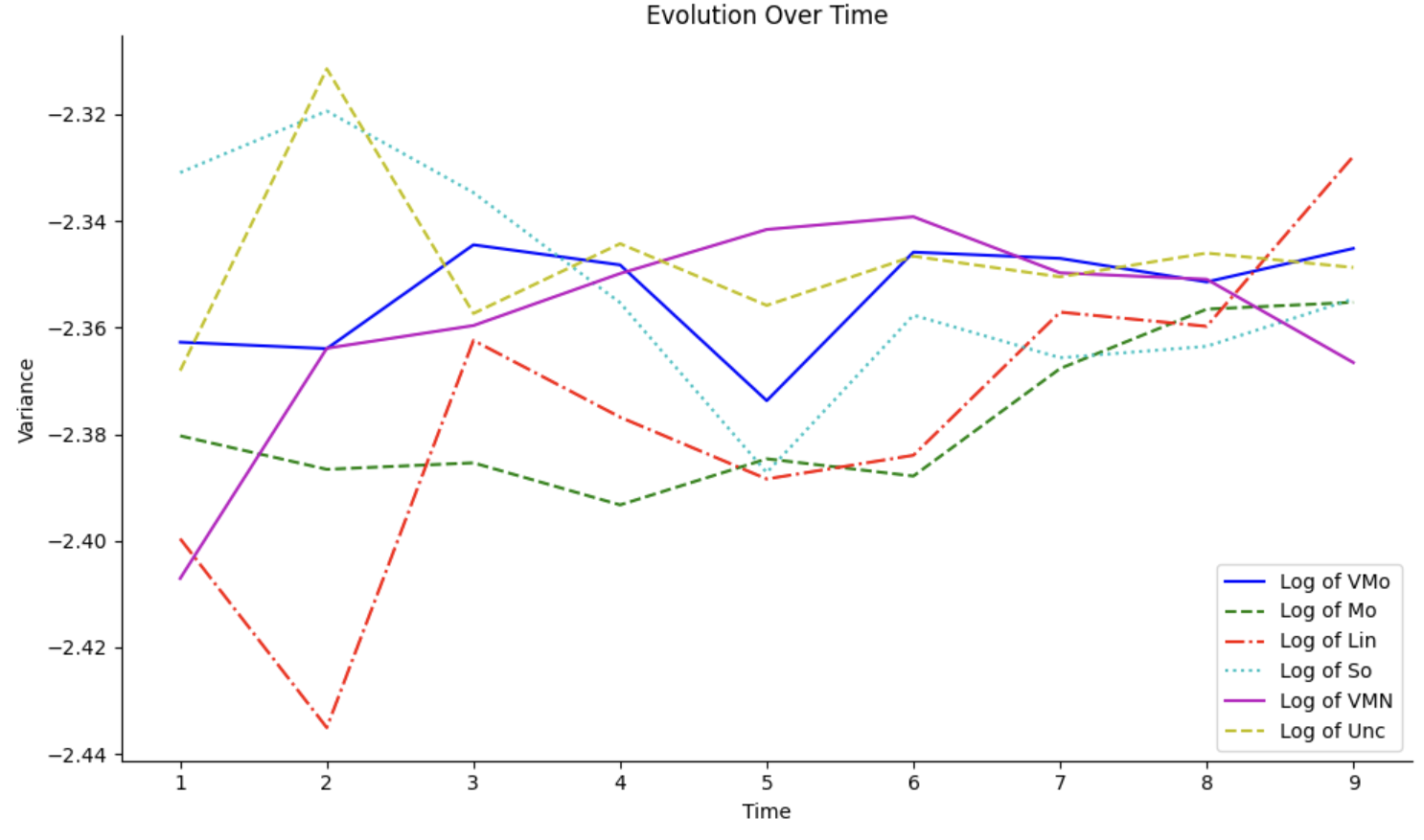}
    \caption{Evolution of Diversity Metrics Across Rounds}
    \label{fig:DiversityEvolutionOneUser}
\end{figure}

\textbf{Diversity Evaluation}
Figure~\ref{fig:DiversityEvolutionOneUser} shows the evolution of volume, ridge leverage scores, and variance over time, based on the sampled item batches across different selection strategies. This visualization evaluates whether the diversity metrics used in the reward function (volume and ridge leverage scores) improve over time and remain consistent with standard statistical measures of diversity, such as variance.

Volume evolution illustrates how the log-volume of the selected item subsets evolves across time steps. The method \texttt{VMN} exhibits consistently higher volume values, indicating that it tends to select more diverse item sets in terms of coverage within the feature space. Conversely, \texttt{Mo} and \texttt{So} show significantly lower volume values, suggesting more redundant or narrowly focused selections. The \texttt{Lin} strategy shows a steady upward trend, reflecting gradual improvement in diversity, although it does not reach the diversity levels of \texttt{VMN} or \texttt{VMo}.

Variance evolution tracks the average variance of item features within the selected batches. While the differences between methods are more subtle here than in volume, \texttt{VMo}, \texttt{VMN}, and \texttt{Unc} consistently maintain slightly higher variance values, supporting the interpretation that these strategies favor more diverse item selections. The \texttt{Mo} and \texttt{So} strategies exhibit relatively low and stable variance throughout, reflecting their tendency to repeatedly select similar items. In contrast, \texttt{Lin} shows a noticeable upward trend toward the final rounds, indicating that although diversity is considered only in the reward function, it gradually explores a wider range of feature space.



Finally, the average ridge leverage scores—an indicator of representativeness and inter-batch diversity—consistently increase over time for all strategies. This trend confirms that as more items are selected, the batches become more representative of the global space. Notably, while \texttt{Unc} converges quickly, indicating effective exploration, \texttt{Mo} shows artificially fast convergence despite poor item coverage, reflecting its tendency to over-exploit a narrow subset of items.

Taken together, these plots validate that the diversity metrics used in the reward function (volume, Risge Leverage Score) are consistent with classical statistical indicators (variance or standard deviation), confirming the reliability of the proposed evaluation approach.

\begin{figure}[H]
    \centering
    \includegraphics[width=0.7\linewidth]{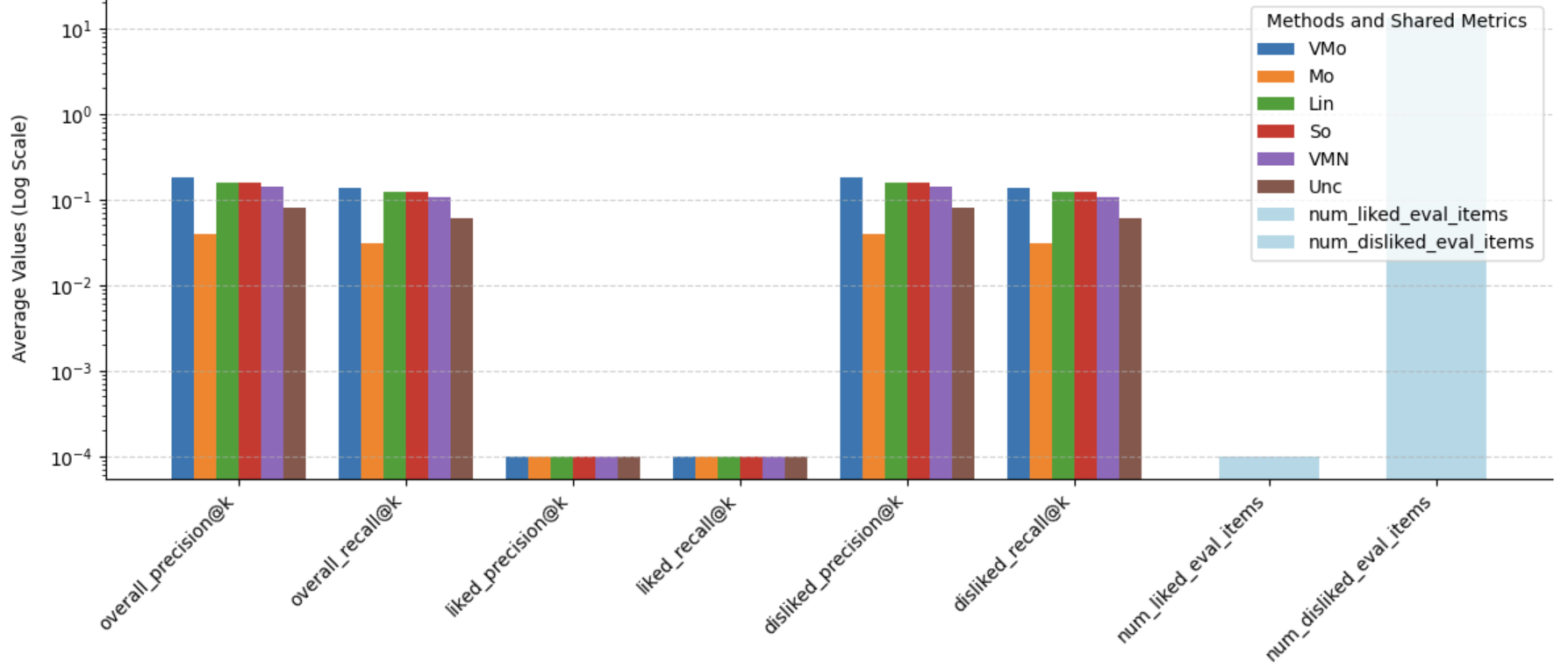}
    \caption{Precision and Recall on Liked and Disliked Items Across Different Selection Strategies}
    \label{fig:relevanceOneUser}
\end{figure}

\textbf{Relevance Evaluation}
Figure~\ref{fig:relevanceOneUser} presents the average performance of different recommendation strategies with respect to relevance-based metrics, computed over a single user. The evaluation includes both aggregate metrics—\emph{overall precision} and \emph{overall recall}—as well as per-category metrics: \emph{liked items precision}, \emph{liked items recall}, \emph{disliked items precision}, and \emph{disliked items recall}. All values are plotted on a logarithmic scale to highlight small differences in low-frequency outcomes.

The left-most bars indicate that the methods \texttt{VMo}, \texttt{Lin}, and \texttt{So} achieve the highest overall precision and recall values, while \texttt{Mo} and \texttt{Unc} lag behind. This strong performance can be attributed to the fact that \texttt{Mo}, \texttt{So}, and \texttt{Lin} all incorporate user-item similarity in their selection formulations, directly favoring items that align closely with user preferences. Notably, \texttt{VMo} consistently outperforms \texttt{VMN}, which focuses solely on diversity. This superior performance is explained by \texttt{VMo}’s multi-objective design: it combines relevance and diversity with equal importance through a dominance-based ranking procedure, ensuring a well-balanced selection at each iteration

The right-most part of the plot shows shared context values: \emph{the number of liked items} and \emph{the number of disliked items}. The large imbalance between these bars confirms that the user has far fewer liked items than disliked ones, which directly affects the attainable precision and recall in the liked-item category. This distribution reinforces the importance of evaluating recommendations not only in aggregate but also per preference category, as it helps expose how different methods manage the trade-off between relevance and noise.

\begin{figure}[H]
    \centering
    \includegraphics[width=0.5\linewidth]{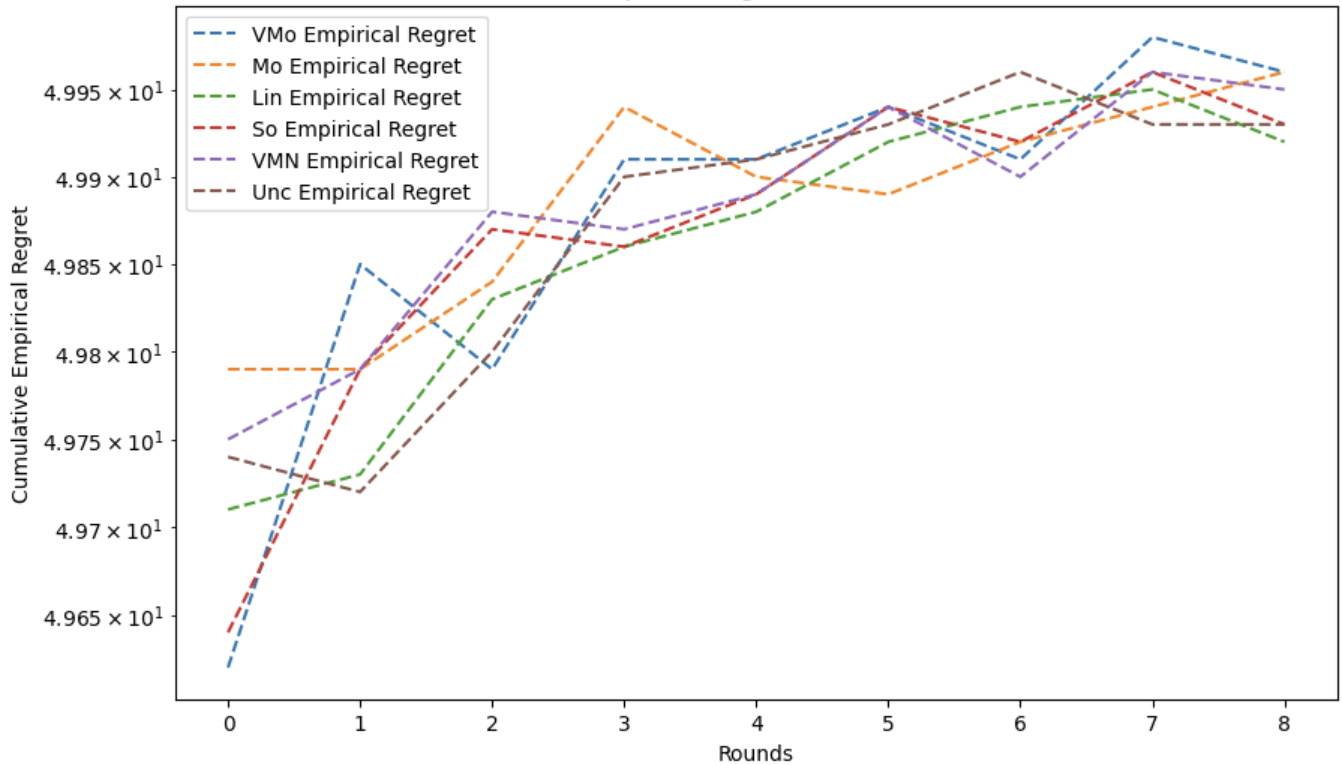}
    \caption{Logarithmic Evolution of Empirical Regret in Terms of Volume}
    \label{fig:learningOneUser}
\end{figure}

\textbf{Learning Evaluation}
Figure~\ref{fig:learningOneUser} presents the cumulative empirical regret over multiple rounds for different selection strategies. Empirical regret quantifies the difference between the reward of an optimal set and that of the selected subset at each round; lower values indicate more effective balancing of exploration and exploitation. 

While user preferences can be inferred from historical data, the added diversity introduced by sequential selections is inherently harder to predict, as it involves a combinatorial item set. Therefore, a relatively high diversity threshold (e.g., a volume of 50) is used as a practical reference to assess whether a selected batch achieves substantial diversity. This offers a reasonable proxy for measuring regret in contexts where diversity is a core objective.

All strategies exhibit stable learning behavior, with cumulative regret values increasing slowly over time. The \texttt{Lin} strategy starts with a lower regret, indicating a good initial performance, but is gradually overtaken by other methods as rounds progress. \texttt{Mo} and \texttt{VMo} show more fluctuating behavior, with \texttt{VMo} ending with the highest regret, implying less efficient learning in later rounds. In contrast, \texttt{Unc}, \texttt{VMN}, and \texttt{So} display smoother regret curves and slightly lower final values. 

These results suggest that methods like \texttt{Unc} and \texttt{VMN} better balance exploration and accuracy over time, maintaining competitive regret while sampling diverse items. This behavior likely stems from the fact that both strategies omit user similarity from their formulations, thereby focusing exclusively on diversity-driven signals. Still, the relatively narrow range of regret across all methods indicates that none achieves a dominant advantage under the experimental conditions considered.

\begin{figure}
    \centering
    \includegraphics[width=0.3\linewidth]{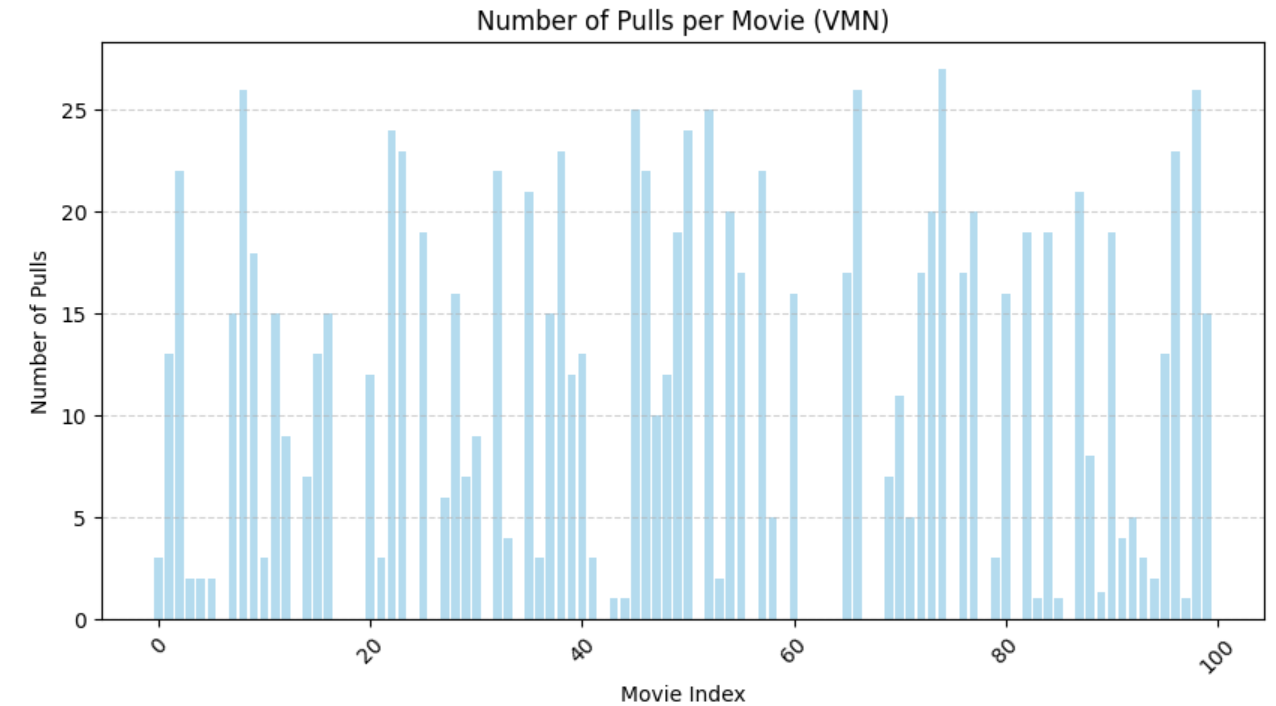}
    \includegraphics[width=0.3\linewidth]{VMN_pulls.png}
    \includegraphics[width=0.3\linewidth]{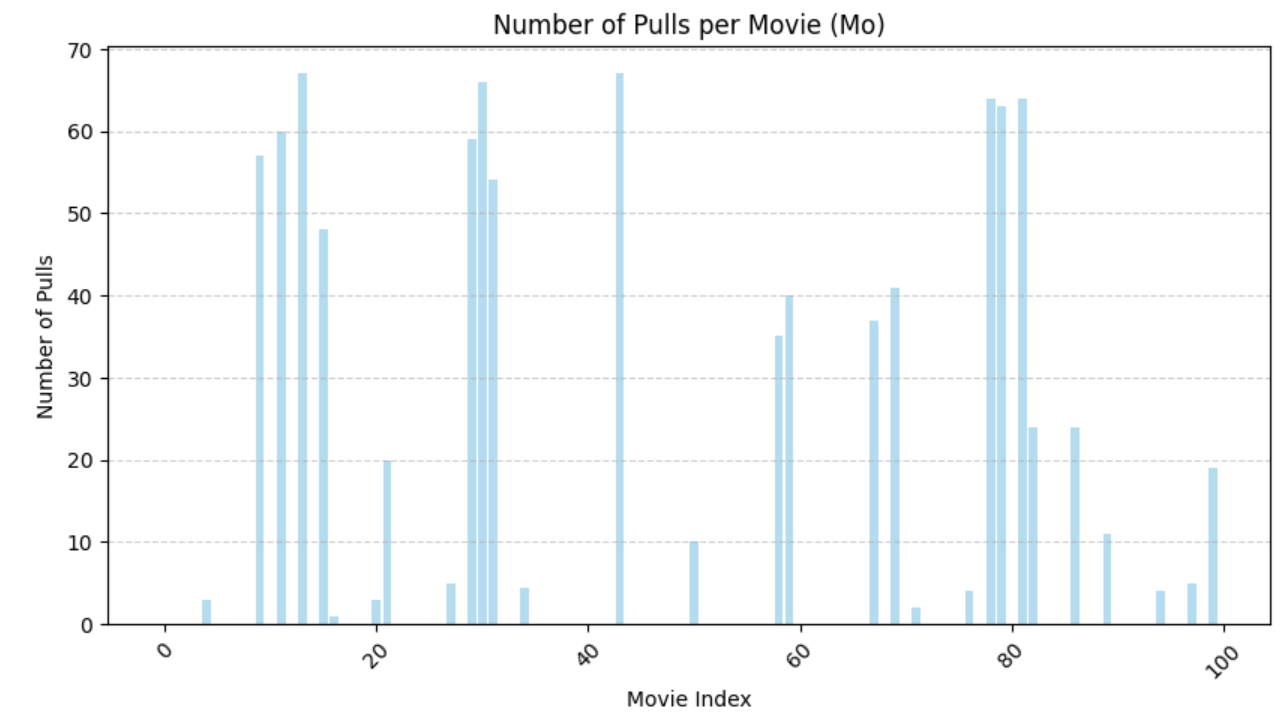}
    \includegraphics[width=0.3\linewidth]{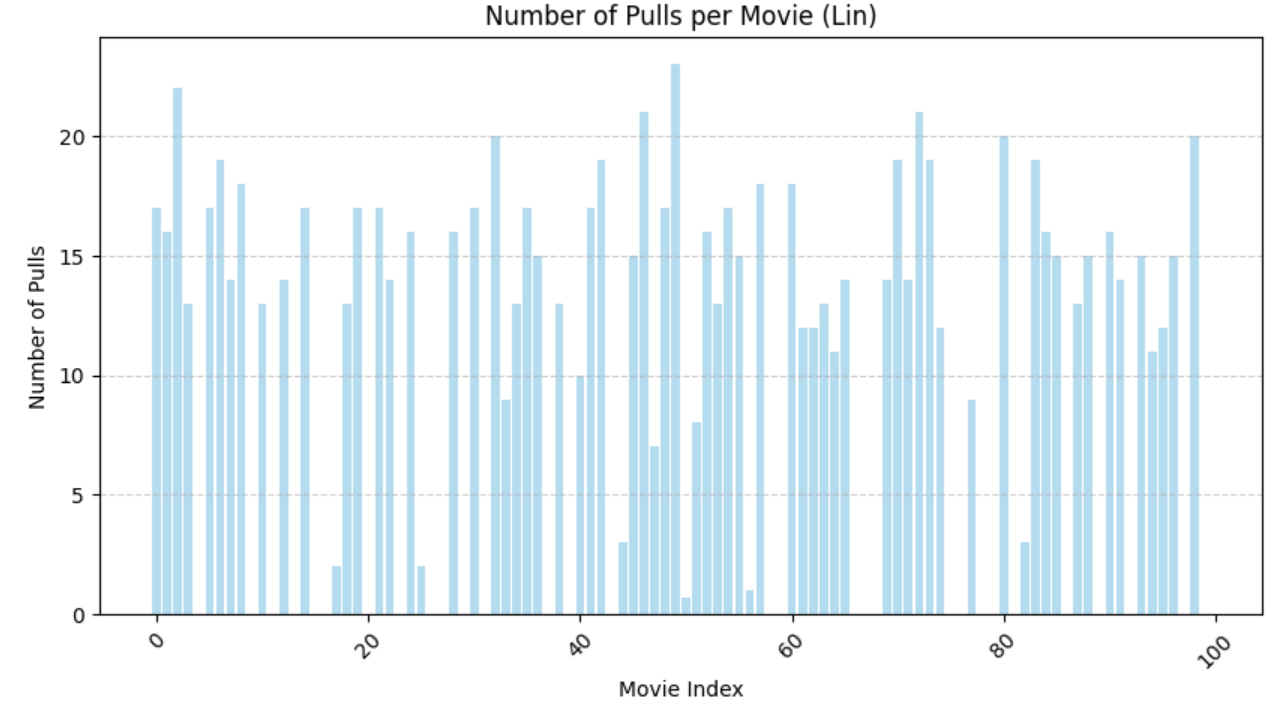}
    \includegraphics[width=0.3\linewidth]{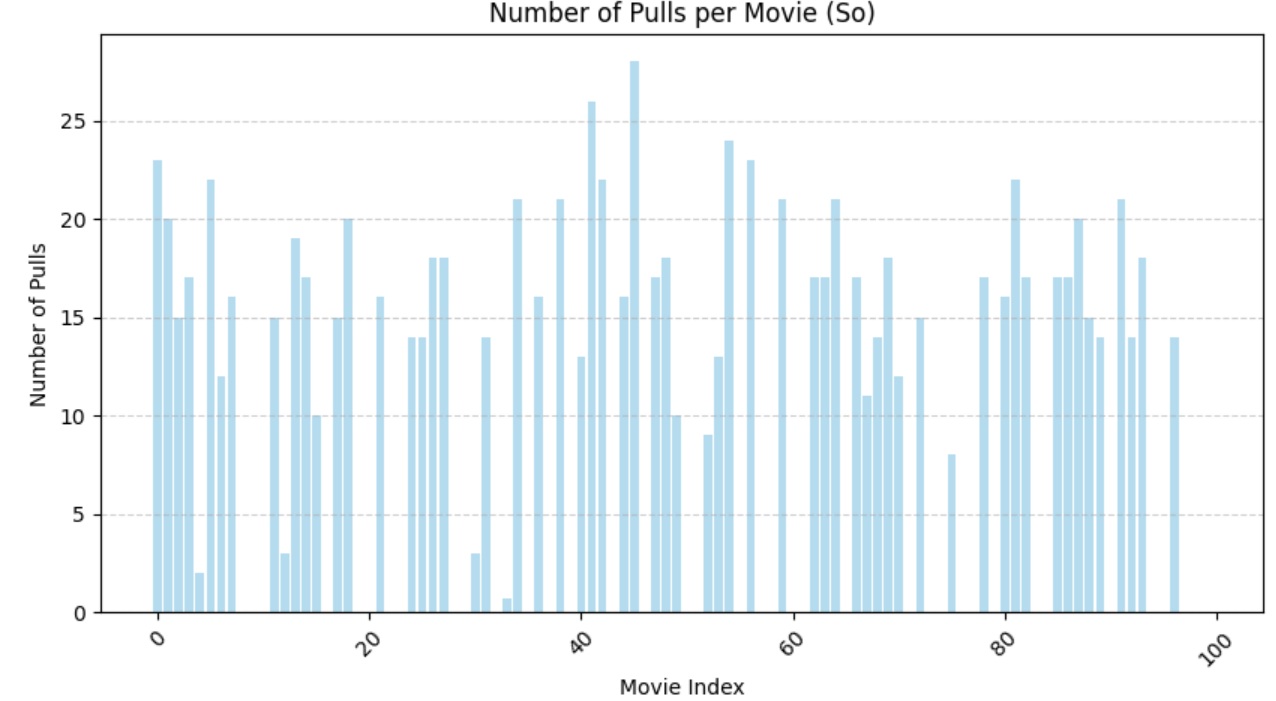}
    \includegraphics[width=0.3\linewidth]{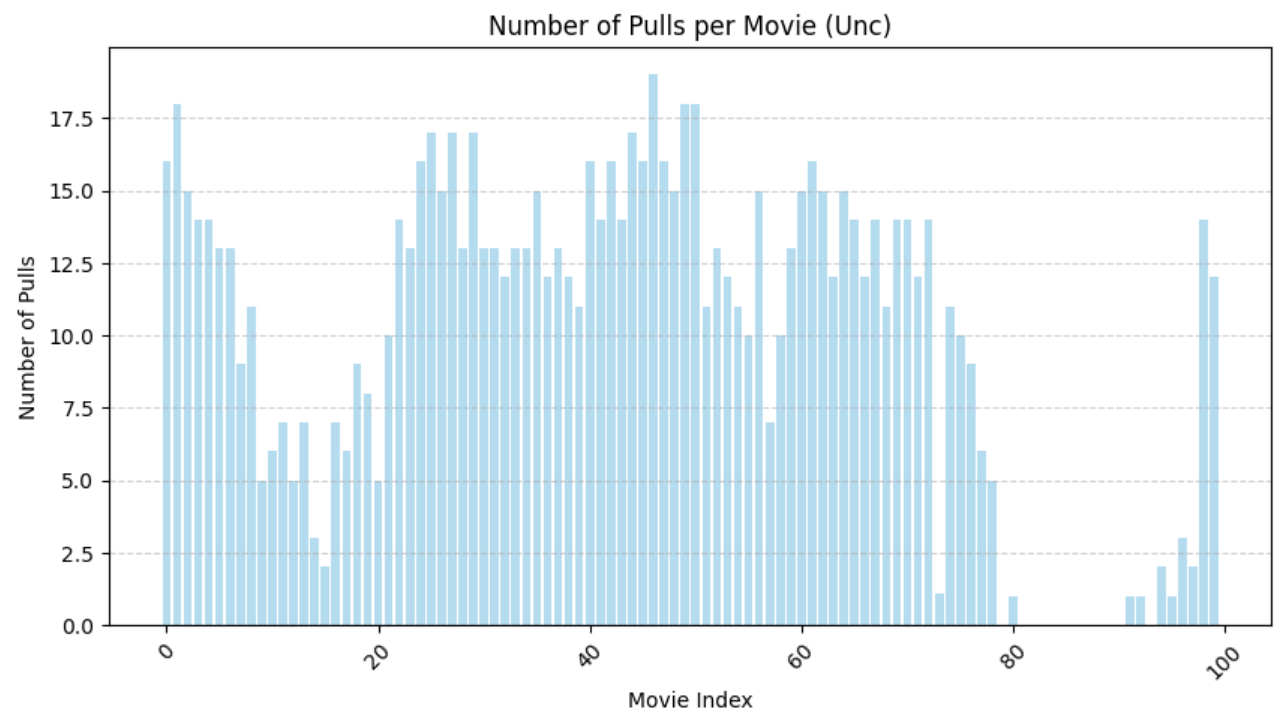}
    \caption{Number of pulls per item under different Selection strategies.}
    \label{fig:Exploration-Coverage}
\end{figure}

\paragraph{Analysis of Item Pull Distribution Across Strategies (exploration)}
Figure~\ref{fig:Exploration-Coverage} displays the number of times each item was selected (or "pulled") under different recommendation strategies. These plots provide insight into the exploration behavior and item coverage of each method.

The \texttt{VMo} and \texttt{Lin} strategies demonstrate a fairly uniform distribution of pulls across the item set, suggesting that both strategies effectively explore a broad range of items. This behavior aligns with their use of diversity-aware signals in the reward function, emphasizing both accuracy and coverage.

In contrast, the \texttt{Mo} strategy exhibits a highly concentrated selection pattern, where only a small subset of items is repeatedly pulled while the majority are ignored. This reflects a strong exploitation bias: the strategy rapidly locks onto high-reward items, potentially optimizing short-term accuracy at the expense of long-term diversity. However, \texttt{Mo} may still be effective in scenarios where items already interacted with are filtered out—such as in some recommender systems—since this constraint naturally prevents repeated exposure and encourages exploration among the remaining items.

The \texttt{So} strategy shows a more balanced pull pattern than \texttt{Mo}, though some concentration around specific items remains. Similarly, \texttt{VMN} achieves wide item coverage, with noticeable variation in frequency.

The \texttt{Unc} strategy, based only on uncertainty-driven reward adjustments, demonstrates a broad and dynamic distribution. While not perfectly uniform, it tends to to favor under-explored areas of the item space, effectively encouraging exploration in uncertain regions while maintaining reasonable attention to accurate items.

Overall, strategies such as \texttt{VMo}, \texttt{Lin}, \texttt{VMN}, and \texttt{Unc} succeed in spreading recommendations across the item space. In contrast, \texttt{Mo} and to a lesser extent \texttt{So}, show limited exploration, favoring exploitation-heavy selection that may not generalize well in the long term.

\section{Discussion}
\label{sec:discussion}
The empirical study conducted in this work offers several insights into the interplay between exploration, diversity, and relevance in sequential recommendation. All strategies leverage a shared uncertainty signal $\theta$, derived from Thompson Sampling based on item pull counts. While the \texttt{Unc} strategy applies $\theta$ directly for ranking, other approaches integrate it multiplicatively with contextual signals—such as user-item and context similarity (History similarity)—to shape more informed selection policies.

The strategies differ notably in their objective formulations (mono- vs. bi-objective) and ranking mechanisms. \texttt{Mo} combines $\theta$ with both user-item and context similarity in an unvectorized bi-objective setting, using dominance-based ranking. \texttt{So}, by contrast, uses only user-item similarity in a mono-objective setup with Top-$K$ selection. \texttt{Lin} blends $\theta$ with both types of similarity via a linear weighted sum, also relying on Top-$K$ ranking. On the other hand, \texttt{VMo} and \texttt{VMN} adopt fully vectorized formulations: \texttt{VMo} applies element-wise multiplication of $\theta$ with both user and context similarity and uses dominance-based ranking, while \texttt{VMN} focuses on context similarity alone and employs Top-$K$ ranking.

Architectural choices in how $\theta$ is combined with relevance and diversity signals strongly influence the observed performance trade-offs. Unvectorized strategies like \texttt{Mo}, and \texttt{So} provide simpler, interpretable baselines that achieve stable short-term regret and faster convergence. However, they tend to focus on a narrower subset of items, particularly \texttt{Mo}, which exhibits a strong exploitation bias when no filtering mechanism is applied. In contrast, vectorized strategies like \texttt{VMo} and \texttt{VMN}, which apply $\theta$ element-wise with contextual similarity, achieve broader exploration and significantly higher diversity.

These vectorized strategies also align better with standard statistical diversity metrics—such as variance and standard deviation—and demonstrate more consistent improvement in intra- and inter-batch diversity over time, accompanied by stable learning dynamics and lower diversity-based regret. The use of Pareto-based dominance ranking in multi-objective settings further enhances their ability to cover the item space effectively, making them well-suited for platforms where novelty, coverage, or discovery are strategic priorities (e.g., cultural or educational domains). Notably, this strategy (VMo) also perform competitively on the relevance objective, demonstrating their capacity to balance diversity with alignment to user preferences.

Overall, the results support the importance of multi-objective formulations in recommendation systems, particularly those that incorporate kernel-based diversity measures, and uncertainty-aware scoring. By flexibly balancing relevance and diversity, such approaches can better align with the nuanced and evolving goals of real-world recommendation platforms.

\section{Conclusion}
\label{sec:conclusion}
This work introduces a \emph{Contextual Diversity-aware Sequential Sampling} framework that integrates diversity-aware rewards into Thompson Sampling to effectively balance exploration and exploitation in recommendation systems. Diversity is captured using a tuned kernel-based metrics—specifically \emph{Volume} and \emph{Ridge Leverage Scores (RLS)}—and incorporated through an advantage ratio formulation, where positive and negative feedback respectively update the Bayesian parameters 
to guide future sampling decisions. By combining this uncertainty signal with user-item and contextual similarity, the framework accommodates both vectorized and unvectorized scoring formulations. Selection is performed via either Top-$K$ or dominance-based ranking, depending on whether these similarity signals are considered jointly, weighted, or individually.


Importantly, no explicit filtering was applied to avoid recommending previously selected items, yet most strategies still explored the item space thoroughly. 
Vectorized strategies—especially \texttt{VMo} and \texttt{VMN}—alongside the unvectorized \texttt{Lin} achieved broader item coverage
, confirming their effectiveness for sustained discovery. 
In contrast, unvectorized methods like \texttt{Mo} and \texttt{So} favored exploitation and exhibited more concentrated selections, making them suitable baselines in stable or relevance-driven environments.


These findings highlight the need for principled reward designs and selection mechanisms aligned with diverse recommendation goals. The proposed kernel-based framework enables detailed analysis of recommendation dynamics and supports robust, adaptive systems. Future extensions may address cold-start settings using only semantic item embeddings and incorporate richer context, such as user budgets, item categories, and temporal preferences, for more personalized recommendations.

\bibliographystyle{plain}  
\bibliography{references}  

\appendix
\end{document}